\documentclass[11pt,leqno]{article}
\usepackage{latexsym}
\usepackage{epsfig}
\usepackage{color}
\usepackage{caption}
\usepackage{subcaption}
\usepackage{framed}

\usepackage{amssymb,amsmath}

\usepackage[noline,nofillcomment,noend,noresetcount,boxed,figure]{algorithm2e}

\numberwithin{equation}{section}

\newtheorem{theorem}{Theorem}[section]
\newtheorem{lemma}[theorem]{Lemma}
\newtheorem{proposition}[theorem]{Proposition}

\newtheorem{corollary}[theorem]{Corollary}
\newtheorem{example}{Example}[section]
\long\def\remark #1{\noindent{\bf Remark:} #1\\}
\def\example #1{\noindent{\bf Example:} #1\\}
\long\def\remarks #1{\noindent{\bf Remarks:} #1\\}
\long\def\claim #1 #2{\bigskip\noindent{\bf Claim {#1}} {\it #2}\bigskip}
\def\xclaim #1 #2{\noindent{\bf Claim {#1}} {\it #2}\bigskip}
\newenvironment{proof}{\noindent{\bf Proof:}}{\hfill $\Box $\\}

\renewcommand{\thetheorem}{\arabic{section}.\arabic{theorem}}

\def\ncas #1 {\noindent {\bf Case #1.}\ }

\def\bipart #1 #2{\bigskip \noindent {\bf #1} {\it #2}}
\def\xbipart #1 #2{\noindent {\bf #1} {\it #2}}
\def\iipart #1 #2{\bigskip \noindent {\it #1} {\it #2}}
\def\xiipart #1 #2{\noindent {\it #1} {\it #2}}
\def\brpart #1 #2{\bigskip \noindent {\bf #1} {#2}}
\def\xbrpart #1 #2{\noindent {\bf #1} {#2}}
\def\irpart #1 #2{\bigskip \noindent {\it #1} {#2}}
\def\xirpart #1 #2{\noindent {\it #1} {#2}}

\def\o {\overline}

\def\mod{{\rm{\ mod\;}}}

\def\case #1{\bigskip\noindent{{\bf Case} {\em #1}:}}
\def\subcase #1{\bigskip\noindent{{\bf Subcase} {\em #1}:}}
\def\numcase #1 #2{\bigskip\noindent{{\bf Case #1} {\em #2}:}}

\def\nclaim #1 {\noindent{\bf Claim #1: }}

\def\obs #1 {\bigskip\noindent{\bf Observation #1: }} 

\def\mathy #1{\ifmmode {#1}\else{$#1$}\fi}
\def\o{\overline}

\begin{document}


\title{Maximum Cardinality $f$-Matching in Time $O(n^{2/3}m)$}

\author{%
Harold N.~Gabow%
\thanks{Department of Computer Science, University of Colorado at Boulder,
Boulder, Colorado 80309-0430, USA. 
E-mail: {\tt hal@cs.colorado.edu} 
}
}


\maketitle
\def\today{\ifcase\month\or
January\or February\or March\or April\or May\or June\or
July\or August\or September\or October\or November\or December\fi
\ \number\day, \number\year}
\def\date#1.#2.{\ifcase#1\or
January\or February\or March\or April\or May\or June\or
July\or August\or September\or October\or November\or December\fi
\ #2, \number\year}
\def\ydate#1.#2.#3.{\ifcase#1\or
January\or February\or March\or April\or May\or June\or
July\or August\or September\or October\or November\or December\fi
\ #2, 199#3}
\def\nydate#1.#2.{\ifcase#1\or
January\or February\or March\or April\or May\or June\or
July\or August\or September\or October\or November\or December\fi
\ #2}
\def\doublespace{\multiply\baselineskip by3\divide\baselineskip by2%
                 \def\doublespace{}}
\def\bigdoublespace{\multiply\baselineskip by2%
                 \def\bigdoublespace{}}
\def\imp{\ifmmode {\ \Longrightarrow \ }\else{$\ \Longrightarrow \ $}\fi}
\def\rimp{\ifmmode {\ \Longleftarrow \ }\else{$\ \Longleftarrow \ $}\fi}
\def\ximp{\ifmmode {\Longrightarrow\ }\else{$\Longrightarrow\ $}\fi}
\def\xrimp{\ifmmode {\Longleftarrow\ }\else{$\Longleftarrow\ $}\fi}
\def\iff{\ifmmode {\ \Longleftrightarrow \ }\else{$\ \Longleftrightarrow \ $}\fi}
\def\xiff{\ifmmode {\Longleftrightarrow\ }\else{$\Longleftrightarrow\ $}\fi}
\def\tru{\ {\bf true}\ }
\def\fal{\ {\bf false}\ }
\def\wrt{\ {\it wrt}\ }
\def\endskip{\medskip}
\def\qed{$\Box$}
\def\qedn{\ \vrule width4pt depth-1pt height7pt }
\def\rqed{\hfill\hbox to 24 pt{\vrule width4pt depth-1pt
height7pt\hfil}\bigskip}
\def\rqedn{\hfill\hbox to 24 pt{\vrule width4pt depth-1pt height7pt\hfil}}
\def\log{\ifmmode \,{ \rm log}\,\else{\it log }\fi}
\def\con {\subseteq}
\def\pcon{\subset}
\def\firstnumstp#1 {\bigskip \noindent{\it Step} #1.\newquad}
\def\numstp#1 {\endskip\noindent{\it Step} #1.\newquad}
\def\newquad{\hskip1ex}
\def\stp#1.{\endskip
\noindent{\it #1 Step.}\newquad}
\def\firststp#1.{\bigskip
\penalty-1000
\noindent{\it #1 Step.}\newquad}
\def\cas#1 {\smallskip\noindent{\bf Case} #1.\ } 
%
%
%
\long\def\sec#1{\bigskip
\penalty-2000%
\noindent{\twelvebf #1}\par\ignorespaces\noindent\ignorespaces}
\def\aorbsec#1{\noindent{\twelvebf #1}}
\def\nsec#1{\penalty-2000%
\noindent{\bf #1\hfill\break}
\hbox to \parindent{\hfill}\ignorespaces}
\long\def\res #1. #2{\bigskip
\penalty-1000
\noindent {\bf #1.}\newquad%
#2 \bigskip}
\long\def\nres #1. #2{\bigskip
\noindent {\bf #1.}\newquad%
#2}
\def\pf{\noindent {\bf Proof.}\newquad}
\def\cont{\ifmmode\star\else$\star$\fi}
\def\+{\tabalign} 
\def\nskp{\def\bigskip{}}
\def\i{($i$) } \def\xi{($i$)}
\def\ii{($ii$) } \def\xii{($ii$)}
\def\iii{($iii$) } \def\xiii{($iii$)}
\def\iv{($iv$) } \def\xiv{($iv$)}
\def\pa{({\it a}) } \def\xpa{({\it a})} 
\def\pb{({\it b}) } \def\xpb{({\it b})}
\def\pc{({\it c}) } \def\xpc{({\it c})}
\def\hi{\hskip20pt\i} \def\hii{\hskip20pt\ii} \def\hiii{\hskip20pt\iii}
\def\ha{\hskip20pt\pa} \def\hb{\hskip20pt\pb} \def\hc{\hskip20pt\pc}
\def\tran{{\buildrel*\over\to}}
\def\n{\rlap{$\>/$}}
\def\({{\rm(}} \def\){{\rm)}}
\def\c#1{\lceil {#1} \rceil}
\def\f#1{\lfloor {#1} \rfloor}
\long\def\boxit#1{\vtop{\hrule
\hbox{\vrule\quad\vtop{\vskip5pt\hbox{#1}\vskip5pt}\quad\vrule}
\hrule}} 
\def\iboxit#1{\vtop{\hrule
\hbox{\vrule\quad\vtop{\vskip5pt\hbox{{\it #1}}\vskip5pt}\quad\vrule}
\hrule}} 
\def\x{\iffalse}
\def\b{\bigskip}
\def\set #1#2{\{ #1:#2 \}}
\def\pset #1#2{( #1:#2 )}
\def\h{\hskip20pt}
\def\hi{\advance\parindent by 20pt}

\def\o{\overline} 
\def\u{\underline}
\def\opn{\hangindent=40pt\hangafter=1}
\def\h{{\hskip 20pt}}
\def\v{\vfill}
\def\hi{\advance \parindent by 20pt}
\def\d{\cdot}
\def \il #1{\log^{(#1)} }
\def\al.{{\it add\_leaf}}
\def\alm{{\it add\_leaf}$\,$}
\def\O{o\hbox{-}smallest}
\def\oS.{\ifmmode{ \o{\cal S} }\else{$\o {\cal S}$}\fi}
\def\oP.{\ifmmode{ \o{\cal P} }\else{$\o {\cal P}$}\fi}
\def\ot.{\mathy{ \o{\cal T} }}
\def\oG{\o G}
\def\oB{\o B}
\def\oE.{\mathy{\overline E}}
\def\p(#1,#2){\ifmmode p(#1,#2) \else{$p(#1,#2)$}\fi}
\def\op(#1,#2){\ifmmode \o{p}(#1,#2) \else{$\o{p}(#1,#2)$}\fi}
\def\lb{\ifmmode \,{ \rm log}_\beta \else{\it log XX }\fi}
\def\wh{\widehat}
\def\wx.{\ifmmode \wh x \else$\wh x$\fi}
\def\wy.{\ifmmode \wh y \else$\wh y$\fi}
\def\wz.{\ifmmode \wh z \else$\wh z$\fi}
\def\wv.{\ifmmode \wh v \else$\wh v$\fi}
\def\Px.{\ifmmode \wh x \else$\wh x$\fi}
\def\Py.{\ifmmode \wh y \else$\wh y$\fi}
\def\Pz.{\ifmmode \wh z \else$\wh z$\fi}
\def\Pv.{\ifmmode \wh v \else$\wh v$\fi}
\def\Pr.{\ifmmode \wh r \else$\wh r$\fi}
\def\Pr.{\ifmmode \wh r \else$\wh r$\fi}
\def\A.{\mathy{{\cal A}}}
\def\B.{\mathy{{\cal B}}}
\def\E.{\ifmmode {{\cal E}}\else{{$\cal E$}}\fi}
\def\F.{\mathy{\cal F}}
\def\H.{\mathy{\cal H}}
\def\M.{\mathy{\cal M}}
\def\P.{\mathy{\cal P}}
\def\Q.{\mathy{\cal Q}}
\def\S.{\ifmmode {{\cal S}}\else{{$\cal S$}}\fi}
\def\T.{\mathy{\cal T}}
\def\mathy #1{\ifmmode {#1}\else{$#1$}\fi}
\def\goin{\hspace{17pt}}

\begin{abstract}
We present an algorithm that finds a maximum cardinality $f$-matching
of a simple graph in time $O(n^{2/3} m)$. Here $f:V\to \mathbb{N}$
is a given function, and an $f$-matching is a subgraph wherein each vertex
$v\in V$ has degree $\le f(v)$.
This result
generalizes a string of algorithms, concentrating on simple bipartite
graphs.  The bipartite case is based on the notion of level graph,
introduced by Dinic for network flow. For general graphs the ``level''
of a vertex is unclear: A given vertex can occur on many different
levels in augmenting trails. In fact there does not seem to be a
unique level graph, our notion of level graph depends on the trails
being analyzed.  Our analysis presents new properties of blossoms of
shortest augmenting trails.

Our algorithm, unmodified, is also efficient on multigraphs,
achieving time $O(\min \{\sqrt {f(V)}, n\}\,m)$,
for $f(V)=\sum_vf(v)$.
\end{abstract}


\def\xmod{{\rm mod\ }}

\def\barj{{\bar j}}

\def\il (#1){\mathy{\ell_i(#1)}}
\def\ol (#1){\mathy{\ell_o(#1)}}
\def\iol (#1){\mathy{\ell_j(#1)}}
\def\iolb (#1){\mathy{\ell_{\barj}(#1)}}
\def\ioL (#1){\mathy{L_j(#1)}}
\def\ioLb (#1){\mathy{L_{\barj}(#1)}}

\def\H#1{\widehat{#1}}

\def\T{TP}
\def\cT{\mathy{\cal T}}

\section {Introduction} 
\label{IntroSec}
\def\ed #1#2{\mathy{\{#1,#2\}}}

The method of shortest augmenting paths leads to many fundamental algorithms
for graphs and networks. Dinic \cite{D} as well as Edmonds and Karp \cite{EK}
introduced the method for maximum network flow.
Its use for matching and small networks was given by
Hopcroft and Karp \cite{HK},  Karzanov \cite{K,K73b}, and
Even and Tarjan \cite{ET}.
These papers show that a maximum cardinality $f$-matching on a bipartite graph
can be found in time
\begin{equation}
  \label{fTimeBoundsEqn}
  \begin{cases}
    O(n^{2/3}\; m)&G \text{ a simple graph}\\
    O(\sqrt {f(V)} \; m)&G \text{ a multigraph.}
\end{cases}
\end{equation}
\noindent
(For definitions of $f$-matching and other terms
please see the Terminology section below.)
These bounds extend to 
problems with edge weights via
cost-scaling,
 in work by Gabow and Tarjan \cite{GT89} 
 (the time increases by a logarithmic factor for scaling).%
 \footnote{The above bound  are the best known for combinatorial algorithms.
Chen and Kyng et.al. \cite{CKLPGS} use interior point methods to achieve time
$m^{1+o(1)}$
with high probability, even for the general problem of
maximum flow in arbitrary networks.}

 Extending these results from bipartite to general graphs is a big jump,
 because of complications caused by
 blossoms, introduced by Edmonds \cite{Ed} in the first efficient
 general graph matching algorithm.
 For general graphs with $f\equiv 1$, i.e., ordinary matching,
 Micali and Vazirani \cite{MV,V1,V2} use the shortest-augmenting-path method
 to achieve time  $O(\sqrt {n} \; m)$. Huang and Pettie \cite{HP}
 achieve time $O(\sqrt{f(V)}\, m\, \alpha(m,n))$ for general multigraphs 
 (using scaling).

 All
 of the above cited algorithms
find shortest augmenting paths
using
the natural, breadth-first-search notion of
 a level graph 
 (\cite[pp. 153-155]{S}, \cite[p.166]{KV},
 \cite[called the ``layered network'', Section 7.5]{AMO}).

 The main contribution of this paper is to extend the simple graph bound
 $O(n^{2/3}m)$ from bipartite to general graphs.
 The difficulty is that the natural notion of level graph
 does not  capture the structure of blossoms 
 needed for this bound. Fundamentally:

 \bigskip
 
 $\bullet$ Vertices no longer have unique levels. An $O(n)$ size graph
 can have $\Theta(n^2)$ different levels.

 $\bullet$  Our level graph is not unique.  It depends on the augmenting trails
 being analyzed.

 $\bullet$ Our analysis uses the natural version of our level graph
 to prove basic properties of blossoms. But we need to remove redundancies
 to deduce the final bound.

 \bigskip
 
 \noindent
 In spite of these difficulties we
 wind up with only two levels for each vertex, just like
 the bipartite case.

 \bigskip
 
 Previous work on simple general graphs was essentially introduced in the break-through
 paper of Duan, He, and Zhang \cite{DHZ}.
 They present a scaling  algorithm for maximum weight
 $f$-matching
  of
simple (general) graphs. It comes within logarithmic factors of the bipartite bound,
$\widetilde{O} (n^{2/3}\, m\, \log W)$ ($W$ the maximum edge weight).
In fact when all weights are 1 this is the first algorithm
to achieve the $n^{2/3}\, m$
bound to within logarithmic factors,
for unweighted $f$-factors of simple general graphs.
(See also \cite{G23b} for 
a algorithm 
on multigraphs with time
$O(\sqrt {f(V) \log f(V)}\,m
\,\log  (f(V) W)$.) 

Our algorithm also runs in time  $O(\min \{\sqrt {f(V)}, n\}\,m)$ on multigraphs.
The first bound is a slight improvement of the above bound of $\cite{HP}$ (and avoids
scaling).

\paragraph{Content and structure of the paper}
We call the new type of vertex levels ``petalevels''. (They are caused by the petals
of blossoms.) We present a family of example graphs that has a quadratic number of petalevels. The construction is not needed to prove our upper bound, but provides motivation.
We also present formulas for ordinary levels and petalevels, without proof but for
more motivation. The formulas can be proved using the approach of \cite{G17} for matching.
In fact our algorithm follows the weighted matching approach to maximum cardinality matching introduced by Gabow \cite{G17}.

The paper is organized as follows.
Section \ref{BackgroundSec} sets the stage.
It starts by reviewing the shortest-augmenting-trail approach
(Section \ref{SATSec})
and the $f$-matching algorithm
(Section \ref{BlossomReviewSec}).
Section \ref{MultPetalevelsSec} presents our example graph with multiple
petalevels.
Section \ref{BoundSec} proves the main result, the $n^{2/3}$ bound
on iterations for simple graphs.
Section \ref{AlgSec} completes our result by providing details of
the matching algorithm.
Appendix \ref{ExampleGraphAppendixSec} completes the description of how
the $f$-matching  algorithm searches the
example graph.

\paragraph*{Terminology and conventions}
We often omit set braces from singleton sets, denoting $\{v\}$ as
$v$. So $S-v$ denotes $S-\{v\}$. We abbreviate expressions $\{v\}\cup S$ to $v+S$.
We use a common summing notation: If $f$ is a function on elements
and $S$ is a set of elements then $f(S)$ denotes $\sum_{s\in S}f(s)$.

The trees in this paper are out-trees (i.e., a root has indegree 0 and all
other nodes have indegree 1).
An {\em out-forest} is a collection of disjoint out-trees.
(We sometimes erroneously
say ``out-tree'' when ``out-forest'' is the proper term; this is done for
convenience and intuition.)
Writing $xy$ for an arc of a tree
implies the arc joins parent $x$ to child $y$.
In this case we also define $\tau(y)$ to be arc $xy$.
(Mnemonically, $\tau$ stands for directed {\em to}.)

We extend parent and child relations to
tree arcs, e.g., arc $xy$ is the parent of $yz$.
A node $x$ is an {\em ancestor}
of a node $y$ allows the possibility $x=y$, unless $x$ is a {\em proper} ancestor
of $y$.

Graphs in this paper are undirected multigraphs.
Loops are allowed (even in simple graphs). A cycle is a connected degree 2 subgraph,  be it a loop,
2 parallel edges, or an undirected graph cycle.
A {\em trail} is a path that is allowed to repeat vertices but not
edges.
For vertices $x,y$ an {\em $xy$-trail} has ends $x$ and $y$.

In  graph $G=(V,E)$ for $S\con V$ and $M\con E$,
$\delta_M(S)$ ($\gamma_M(S)$)
denotes the set of edges of $M$ with exactly one
(respectively two) endpoints in $S$.
We sometimes write 
For a set of vertices $S\subseteq V$ and a subgraph $H$ of $G$,
$\delta(S,M)$ ($\gamma(S,M)$).
$d_H(v)$ ($d(v,H)$) denotes the degree of vertex $v$ in $H$.  
When
referring to the given graph $G$ we often omit the last argument and write, e.g.,
$\delta(S)$. 
(For example a vertex $v$ has $d(v)=|\delta(v)|+2|\gamma(v)|$.)

We omit $M$ (writing $\delta(S)$ or $\gamma(S)$) when
$M=E$. A loop at $v\in S$ belongs to
$\gamma(S)-\delta(S)$.
For multigraphs $G$ all of these sets are
multisets.

For an undirected multigraph $G=(V,E)$ with function $f:V\to \mathbb Z_+$,
an {\em $f$-matching} is a subgraph where each vertex $v\in V$ has
degree at most $f(v)$.
A maximum (cardinality) {\em $f$-matching} contains the greatest possible
number of edges.
Vertex $v$ is {\em free} if strict inequality holds.

We often call an $f$-matching a {\em matching}.
We refer to an ordinary matching as a 
{\em 1-matching} (i.e., $f$ is 
identically 1 for all vertices).
$def(x)$ is the deficiency of vertex $x$ in the current matching $M$,
$def(x)= f(x)- |\delta(x,M)|-2|\gamma(x,M)|$.

Consider a graph $G$ with an $f$-matching $M$.
An {\em augmenting trail} is an alternating trail $A$ that
begins and ends at a free vertex,
such that $M\oplus A$ is
a valid matching, i.e., the two ends of the trail still satisfy
their degree bound $f$. (The trail may be closed,
i.e., $A$ begins and ends at the same vertex. {\em Alternating} means
consecutive edges of $A$ have opposite M-types.)

Additional notation for matchings is given at the start of
Section \ref{BlossomReviewSec}.

\def\bj{{{ \boldsymbol{\overline} } j}}
\def\bj{{\overline j}}
\def\barj{{\bar j}}

\section{Background}
\label{BackgroundSec}
\subsection{Shortest Augmenting Trails}
\label{SATSec}
Consider an $f$-matching $M$.
Let $S$ be an augmenting trail for $M$. $|S|$ denotes the length of
$S$. $S$ is a {\em shortest augmenting trail} if it has the minimum length
possible. We call $S$ a {\em sat}.

Hopcroft and Karp \cite{HK} and Karzanov \cite{K} proved the
fundamental result that for 1-matching the length of a sat is
nondecreasing.  The proof immediately extends to $f$-matching.  For
completeness we sketch it.

\begin{lemma}
\label{HKLemma}
Let $M$ be an arbitrary $f$-matching. Let $S_1$ be a sat
  for $M$ and $S_2$ a sat for $M\oplus S_1$.
  Then $|S_1|\le |S_2|$.
\end{lemma}

\begin{proof}
  Let $N$ be the $f$-matching resulting from augmenting the two sats,
$N=(M\oplus S_1)\oplus S_2$.
Since $M\oplus N = S_1\oplus S_2$
    we have \[|M\oplus N|\le |S_1\cup S_2| \le |S_1|+|S_2|.\]
We claim 
  $M\oplus N$ contains 
  two edge-disjoint trails $T_1,T_2$ that are augmenting for $M$.
This implies
  \[|M\oplus N| \ge  |T_1| +|T_2| \ge 2|S_1|.\] 
    Combining these two inequalities gives  $|S_1|+|S_2|\ge 2|S_1|$,
     $|S_2|\ge |S_1|$ as desired.
So to complete the proof we need only prove the claim.

For $i=1,2$ let trail $S_i$ have ends $v_i$ and $w_i$.
(These four ends need not be distinct, in fact all four may be
the same vertex.) Enlarge $M$ to an $f$-matching $M'$ by adding
artificial edges $v_iw_i$, $i=1,2$. Every vertex $u\in V$
has $d(u,N)=d(u,M')$. So $N\oplus M'$ is a collection of edge-disjoint
alternating circuits. Removing the two artificial edges creates
two alternating trails that begin and end
 with edges of $N-M$ and hence are augmenting for
 $M$. (In detail let $C_i$ be the circuit containing $v_iw_i$, $i=1,2$.
 If $C_1\ne C_2$ then the desired augmenting trails are
 $T_i= C_i -v_iw_i$. If $C_1= C_2$ then $C_1-v_1w_1-v_2w_2$
 is the disjoint union of trails $T_1$ and $T_2$.)
\end{proof} 

The lemma leads to the following  high-level algorithm for finding a maximum cardinality $f$-matching.
Define a  {\em blocking set} to be a
maximal size collection of edge-disjoint shortest augmenting trails.
The algorithm repeatedly finds
a blocking set and augments each of its trails. When no augmenting trail
exists the matching is maximum.

Our goal is to show
a maximum cardinality $f$-matching on a general graph
can be found in time \eqref{fTimeBoundsEqn}.
Section \ref{AlgSec} shows a blocking set can be found in time $O(m)$.
So the goal is achieved 
by showing the high-level algorithm finds 
$O(\min \{\sqrt {f(V)}, n\}$ blocking sets on multigraphs and
$O(n^{2/3})$ blocking sets on simple graphs.
The arguments are as follows.

Let $M$ be the $f$-matching at some point in the high-level algorithm.
Let $s$ be the length of a sat for $M$.
We start with the simple
analysis for multigraphs.

\paragraph{Proof of the \boldmath$\sqrt {f(V)}$ bound}
A maximum cardinality matching $M^*$ gives a collection of
$(|M^*|-|M|)/2$ disjoint augmenting trails.  Each such trail is
alternating. Its length is $\ge s$.  So it has $\ge (s-1)/2$ matched
edges. This imples there are $\le (f(V)/2) / (s-1)/2$ trails.
When $s= \sqrt
{f(V)}+1$ there are $\le \sqrt {f(V)}$ such trails.  So the number of
searches in the algorithm is $\le 2\sqrt {f(V)}+1$.

\bigskip

\paragraph{Proof of the \boldmath $n$ bound}
Any vertex $v$ occurs at most twice in a sat.  In proof, if a sat
enters $v$ on edge $uv$ and reenters $v$ on $u'v$, then the two edges
have opposite M-types (otherwise we can omit the subtrail from $uv$ to
$u'v$.)  Thus the algorithm halts with $s\le 2n$.

\bigskip


\paragraph{Proof of the \boldmath$n^{2/3}$ bound, high level}
The key property is

\bigskip

\noindent $(*)$ \hskip60pt
Any $f$-matching has at most $4(n/s)^2$ more edges than $M$.

\bigskip

\noindent
Taking $s=2^{2/3}n^{2/3}$ shows after $\le 2^{2/3}n^{2/3}$ searches, at most
$(4/2^{4/3}) n^{2/3}=2^{2/3}n^{2/3}$ augmenting trails need be found.
So the total number of searches is at most $2^{5/3}n^{2/3}<4n^{2/3}$.

\bigskip
Section \ref{BoundSec} will prove $(*)$.
We do this by defining a level graph $LG$ and identifying a
bottleneck level, as in \cite{ET} and \cite{K}.
Ultimately the argument is similar to the bipartite argument:
We show some layer of $LG$ has 
$\le 4(n/s)$ nodes, hence  $\le 4(n/s)^2$ edges.

Our approach, both for proving $(*)$ and for
our corresponding efficient algorithm, is
via weighted matching, as presented for ordinary matching by Gabow \cite{G17}.
In detail suppose we seek a sat for a given $f$-matching $M$.
Assign weights to the edges of $G$ by
\begin{equation}
\label{wDefnEqn}
w(e)=
\begin{cases}
0 & e \notin M\\
2 & e\in M.
\end{cases}
\end{equation}
The (incremental) weight of an alternating trail $T$ is defined as 
\[w(T,M)= w(T-M) -w(T\cap M).\]
An alternating trail $T$ that starts 
with an unmatched edge emanating from a free vertex
has length
\begin{equation}
\label{LengthEqn}
|T|=-w(T,M)+|T|\xmod 2.
\end{equation}
So a sat is an augmenting trail with the greatest possible weight.
Edmonds' algorithm for ordinary matching, as extended 
to $f$-matchings by Gabow \cite{G18}, finds such a trail.

Combining this idea with the high-level algorithm
of Hopcroft and Karp gives the pseudocode
for our algorithm in Fig. \ref{BlockingAlgFig}.
We fill in the details of this algorithm
in Section \ref{AlgSec}.
First we will use this outline to
prove $(*)$, in Section \ref{BoundSec}.

\def\myif #1{{\sf {if}} (#1)}
\def\myelif #1{{\sf else if} (#1)}
\def\myifel #1 #2 #3 {{\sf if} (#1) {{#2}} {\sf else} {{#3}}}
\def\myelse {{\sf else}}

\def\myfor #1{{\sf for} (#1)} 
\def\mywhile #1{{\sf while} (#1)}
\def\myloop {{\sf loop}}

\long\def\mycom #1{{\tt /* {\hskip-3pt#1\hskip-3pt} */}}

\def\myand {{\sf and }}
\def\myor {{\sf or }}
\def\myto {{\sf to }}
\def\myreturn{{\sf return}}
\def\mybreak{{\sf break}}
\def\myhalt{{\sf halt}}
\def\sm.{\mathy{\cal S^-}}
\def\myproc{{\sf procedure}}
\newlength{\Efigwidth}
\newlength{\vmargin}

\setlength{\Efigwidth}{\linewidth}
\setlength{\vmargin}{15pt}

\begin{figure}[t]
\begin{framed} 
\begin{minipage}[t]{\Efigwidth}

\vspace{\vmargin}
\setlength{\parindent}{50pt}


Initialize $BlockBound$ to $\min \{\sqrt {f(V)}, 2n\}$ for multigraphs
or $n^{2/3}$ for simple

graphs. Then execute the following:

\bigskip

$M \gets $ a maximal $f$-matching, $\Delta\gets 0$

\mywhile {$\Delta< BlockBound$}

{\hi 


\myfor {every edge $e$} $w(e)\gets $ \myifel {$e\in M$} {$2$} {$0$}
 
\myfor {every vertex $v$} $y(v)\gets 1$ 

execute an $f$-matching  search 
to find a maximum weight augmenting trail

\myif {no augmenting trail is found}  {\myreturn\ $M$}

$\B. \gets $ a blocking trail set

augment the trails of \B.

$\Delta\gets \Delta+2$

} 

\bigskip

\myloop

{\hi 

search for an augmenting trail $T$

\myif {no augmenting trail is found}  {\myreturn\ $M$}

augment $T$
}

\vspace{\vmargin}

\end{minipage}
\end{framed}

  \caption{Blocking trail algorithm to find a maximum
  $f$-matching.}
\label{BlockingAlgFig}
\end{figure}

\subsection{Weighted Blossoms}
\label{BlossomReviewSec}
\begin{figure}[t]
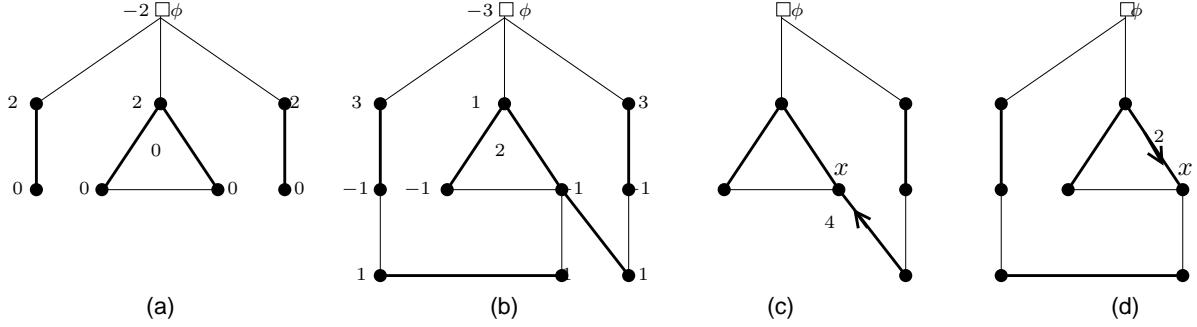

\centering
\input SimplePetal.pstex_t
\caption{Blossoms and petalevels:
(a) Search structure for $y(\phi)=-2$. Vertex labels are $y$ values,
and the triangle blossom has $z$-value 0.
Edge weights are given by \eqref{wDefnEqn}.
(b) Search structure for $y(\phi)=-3$, with $y$ and $z$ values.
One blossom $B$ contains all vertices and has $z$-value 0.
$B$ can be parsed using the subblossom of (c) or of (d). Vertex $x$, as an outer
vertex, is on level 2 in (d) and level 4 in (c).}
\label{SimplePetalFig}
\end{figure}

The proof of $(*)$ requires some basic properties of $f$-matching
blossoms. We present them here. Readers familiar with \cite{G18}
need only skim this section to review the notation.
Fig. \ref{SimplePetalFig} illustrates
the discussion.

We begin with more terminology for matchings.

Contracting a subgraph 
contracted vertex (all internal edges including internal loops
disappear).  We will even contract subgraphs that just contain a loop.
We use the following notation for contractions.  Let $\o G$ be a graph
derived from $G$ by contracting a number of vertex-disjoint subgraphs.
$V$ ($\o V$) denotes the vertex set of $G$ ($\o G$), respectively.  A
vertex of $\o G$ that belongs to $V$ (i.e., it is not in a contracted
subgraph) is an {\em atom}.  We identify an edge of $\o G$ with its
corresponding edge in $G$. Thus an edge of $\o G$ incident to a
contracted vertex is denoted as $xy$, where $x$ and $y$ are
$V$-vertices in distinct $\o V$-vertices, and $xy\in E$.  Let $H$ be a
subgraph of $\o G$. The {\em preimage} of $H$ is a subgraph of $G$
consisting of the edges of $H$, plus the subgraphs whose contractions
are vertices of $H$, plus the atoms of $H$.  $\o V(H)$ ($V(H)$)
denotes the vertex set of $H$ (the preimage of $H$), respectively.
Similarly $\o E(H)$ ($E(H)$) denotes the edge set of $H$ (the preimage
of $H$), respectively.

Figures in this paper use the following conventions.
Matched edges are drawn heavy, unmatched edges light.
Free vertices are drawn as rectangles.
Edge weights are always given by \eqref{wDefnEqn}.
Trails are indicated by arrowheads on their edges.

When discussing a matching $M$, 
the {\em M-type} of an edge $e$ is $M$ or $\o M$ depending on whether
$e$ is matched or unmatched, respectively.
We usually denote an arbitrary M-type as $\mu$,
and $\mu(e)$ denotes the M-type of edge $e$.

Let $G$ be a graph with an $f$-matching.
A trail is {\em closed} if it starts and ends at the same vertex.
A trail is {\em alternating} if  every 2 consecutive edges have opposite M-types.
The first and last edges of a closed trail are not consecutive.

Blossoms are defined recursively as follows.  Let $\o G$ be a graph
derived from $G$ by contracting a family \A. of zero or more
vertex-disjoint blossoms.  Let $C$ be a closed trail in $\o G$ that
starts and ends at a vertex $\alpha\in V(\o G)$.  The preimage of $C$
is a {\em blossom} $B$ with {\em base vertex} $\beta(B)$ if $C$ has
the following properties:

\bigskip


If $\alpha$ is an atom then $C$ starts and ends with edges of the same
$M$-type.  $B$ has this $M$-type and $\beta(B)=\alpha$.

If $\alpha\in \A.$ then
$B$ has the same $M$-type as $\alpha$ and
$\beta(B)=\beta(\alpha)$.

\bigskip

If $v$ is an atom of $C$ then every 2 consecutive edges of $\delta(v,
C) $ alternate.

If $v\in \A.\cap C$ then $d(v,C)=2$.
Furthermore
if $v\ne \alpha$ then
$\delta(\beta(v),C)$ contains an edge of  opposite $M$-type from $v$.

\bigskip

Note that as a subgraph of $G$, a blossom $B$ has vertex and edge sets
$V(B)$ and $E(B)$. Also note
in Fig.\ref{SimplePetalFig} the blossom of part (b)
can be viewed as being formed from the blossom of part (c)
or the blossom of part (d). (This nonuniqueness is usually irrelevant.)

In addition we define
the {\em base edge} of a blossom $A$, denoted $\eta(A)$.
$\eta(A)$ is an edge that leaves $V(A)$ from vertex $\beta(A)$,
and it has opposite $M$-type from $A$.
Every blossom with base $\beta(A)$ has the same base edge $\eta(A)$.
So now assume $A$ is the maximal blossom with base vertex $\beta(A)$.
If $A$ is in the closed trail
$C$ of some blossom, $\eta(A)$ is the edge required in the above definition.
(In case both edges of $\delta(\beta(v),C)$ qualify, the choice for $\eta(A)$ is arbitrary.)
If $A$ is a maximal blossom (i.e., not in such a closed trail)
$\eta(A)$ is some chosen edge (of opposite $M$-type from $A$); it will be clear from context. 

Blossoms where $\beta(A)$ is a free vertex are a 
special case. We say the blossom is free. A free blossom
has $M$-type $\o M$ (since an augmenting trail always begins
with an unmatched edge). Also the vertex $\beta(A)$ has deficiency 1. (Otherwise
the edges of $A$ contain an augmenting trail, and the blossom structure
is irrelevant.) For a 
free blossom $\eta(A)$ 
is a matched edge from an artificial vertex $\varepsilon$ to $\beta(A)$.

For any blossom $B$, any edge leaving $B$ other than $\eta(B)$ is a
{\em petal} of $B$.
Blossom $B$ is {\em heavy} ({\em light}) if its M-type is $M$ ($\o M$), respectively.

In a blossom $B$ with base vertex $\beta$,
each vertex
$v\in V(B)$ has 2 associated  $v\beta$-trails
in $E(B)$,
$P_i(v,\beta)$, $i=0,1$. Each $P_i$
is alternating and
ends with an edge whose $M$-type is that of $B$.
The starting edge for  $P_1(v,\beta)$
has the same $M$-type as $B$; it has  the opposite $M$-type for
$P_0(v,\beta)$. There is one exceptional trail: $P_0(\beta,\beta)$
is the length 0 trail $\beta$.

Next we describe the linear program dual variables for weighted matching,
as they are maintained in \cite{G18} (extending \cite{E}).
Assume every edge of $G$ has a numerical {\em weight} $w(e)$. 
We define a subset of the edges incident to a blossom $B$ by
\[I(B)= \delta(V(B),M)\oplus \eta(B).\]
(Here $M$ is the matching and $I$ stands for ``incident''.)
There are 
two dual functions
$y:V \to \mathbb {R}$,
$z:2^V \to \mathbb {R_+}$.
Define
$\H{yz}:E\to \mathbb {R}$ by
\begin{equation}
\label{fHyzEqn}
\H{yz}(e) = y(e)  + z\set {B} {e \in \gamma(B)\cup I(B)}.
\end{equation}
The duals satisfy the LP feasibility conditions, which (using CS for blossoms)
are:
\begin{eqnarray*}
\text{$e$ matched}&\Rightarrow&
\text{$e$ is {\em underrated}, i.e., $\H{yz}(e) \le w(e)$}\\
\text{$e$ unmatched} &\Rightarrow&  \text{$e$ is {\em dominated}, i.e.,
$\H{yz}(e)$ is  $\ge w(e)$}
\end{eqnarray*}
We say $e$ is {\em tight} if equality holds, otherwise {\em loose}.
The terms {\em strictly dominated} and {\em strictly underrated}
refer to the possibilities  $>w(e)$ and $< w(e)$ respectively.

A {\em positive blossom} has positive $z$-value. 
For any positive blossom $B$ every edge of $E(B)+ \eta(B)$
is tight. 
(A {\em zero blossom} 
has $z(B)=0$. Note that
the subgraph of a zero blossom can be
part of the subgraph
of a positive blossom.
For example 
the blossoms of Fig.\ref{SimplePetalFig}(b)--(d) are zero blossoms.
We can make (b) a positive blossom (lower $y(\phi)$ to $-4$)
but (c) or (d) remains a zero blossom.)


The last property of duals 
is that all free vertices $\phi$ have the same
$y$-value $y(\phi)$.
A {\em structured matching} consists of a matching, duals $y,z$,
and positive blossoms 
wherein every 
unmatched (matched) edge is  dominated (underrated) respectively;
every edge in a positive blossom subgraph is tight; and
every free vertex has the same $y$-value.

\bigskip

Edmonds' algorithm and its extension to $f$-matching
both work by constructing a
``search structure'' \S..
The structure consists of atoms and blossoms, and \oS. denotes
\S. with all blossoms contracted.
\oS. 
is a forest, and we call the elements of $V(\oS.)$ ``nodes''.
(Recall our tree terminology from Section \ref{IntroSec}.)
Every edge of \S. is tight.
The roots of \oS. are the free vertices/blossoms.
If $v$ is an atom of \oS.
then $\tau(v)$ alternates with each child edge of $v$.
(For a free vertex $v$ take
$\tau(v)=\varepsilon v$.) The base edge of a blossom $B$ in \S.
is its parent edge $\tau(B)$.

Every node of \oS. has a corresponding path to the root of its
search tree. These paths expand to trails in $G$ for every
vertex of \S.. We denote the trails as $P_j(v)$, defined as follows.
Call an atom $v$ of \oS. 
{\em inner} if $\tau(v)$ is unmatched and
{\em outer} if $\tau(v)$ is matched.
Correspondingly an {\em io-type} is $i$ or $o$,
for inner and outer respectively.
A vertex $v$ in a blossom has both io-types.
Any vertex of \S. has a trail $P_j(v)$. Here
$j$ is the io-type of $v$.
$P_j(v)$  starts with an edge from $v$ that is
unmatched (matched) for $j=i$ ($j=o$).
It ends on an unmatched edge to a free vertex.
$P_j(v)$ starts with $\tau(v)$ if $v$ is atomic.
If $v$ is in a blossom $P_j(v)$
starts with
the trail $P_i(v)$, where $i$ is chosen
to correspond with $j$.
A node on a path to the root in \oS. expands in a $P_\cdot$ trail
in the obvious way.
Every trail $P_j(v)$ is alternating.

For a blossom $B$, two edges of $\delta(B)$
{\em alternate} if one of the edges is $\eta(B)$.
The motivation for this definition is that the two edges
can be combined with the appropriate $P_i$ trail to form
an alternating trail in $G$. Note that with this definition,
$\oS.$ is an alternating forest, i.e., its edges alternate at every
node. 

$P_j(v)$ is a shortest alternating trail that starts with an
unmatched (matched) edge for $j=i$ ($j=0$)
and ends with an unmatched edge to  a free vertex.
(This can  be proved following the development in
\cite{G17} for 1-matching, or using the tracking algorithm of
Section \ref{BoundSec}. But we do not use this fact.)
We denote the length of $P_j(v)$ as $l_j(v)$.
$l_j(v)$ is the basic level of a vertex in the level graph. 
But unlike level graphs of previous papers,
vertices have other levels. We call these other levels {\em petalevels},
since they are caused by petals in shortest trails.%
\footnote{In more detail ordinary levels are formed by trails that
enter every positive
blossom on its base edge. So they are ``basic levels''. In contrast
petalevels enter at least one positive blossom on a petal.}
Fig. \ref{SimplePetalFig}(c)-(d) show a vertex with ordinary level
$l_o(x)=2$ and petalevel 4.

\begin{figure}[t]
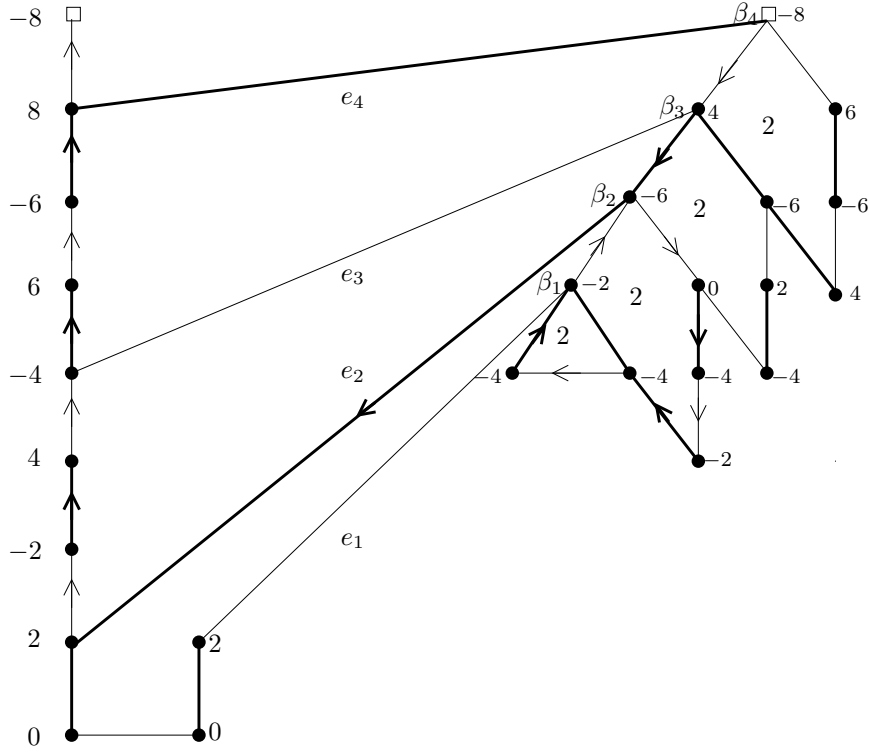

\centering
\input RecPetal.pstex_t
\caption{Vertices have multiple petalevels in graph $EG(4)$.
Four sats of length 17 are formed by ``cross edges'' $e_i$, $i=1,\ldots,4$.
Arrows give $e_2$'s sat.
Vertex $\beta_i$, $i=1,\ldots 4$, is the base of blossom $B_i$,
which consists of the edges touching the $i$ lowest 2's ($B_4$
consists of all the edges descending from $\beta_4$,
and $B_i, i<4$ consists of all but the rightmost edge descending from
$\beta_i$).
The base vertices  occur
at various levels in the sats: The table gives the level $l(B_j)$
of base $\beta_j$ in the sat for $e_i$.
}
\label{RecPetalFig}

\vspace{10pt}

\begin{tabular}{|c|c|c|c|c|}\hline
&$l(\beta_1)$&$l(\beta_2)$&$l(\beta_3)$&$l(\beta_4)$\\ \hline
$e_1$&6&&&\\ \hline
$e_2$&8&9&& \\ \hline
$e_3$&10&11&12&\\ \hline
$e_4$&12&13&14&15\\ \hline
\end{tabular}

\end{figure}


\subsection{Multiple Petalevels}
\label{MultPetalevelsSec}
This section provides a central motivation for our proof of $(*)$,
an example graph
with too many petalevels
to include in the desired level graph.
Strictly speaking the example is not needed to prove $(*)$.
But we include it here
to provide an in-depth look at the 
search structure \S., to illustrate
how multiple petalevels arise,
and to provide
details about how
the example graph's search structure can actually occur in our
algorithm
Fig. \ref{BlockingAlgFig}. (The remaining details for this are given in
Appendix \ref{ExampleGraphAppendixSec}.)

The example graph $EG(b)$ 
has $b$ blossoms, for any even $b \ge 2$. $EG(b)$ has
$\Theta(b)$ vertices and edges.
It has sat length $4b+1$ and importantly, 
$\Theta(b^2)$
petalevels. Fig.\ref{RecPetalFig}
gives a simple example, with $b=4$ and
sat length 17. 
Its table
shows there are
6 petalevels (that are not ordinary levels).

As usual vertical level in the figure corresponds to level, i.e., shortest
distance from a free vertex. So the two vertices with $y$-value 0 are at level
8. Edges drawn vertically are in the search structure \S..

The blossoms $B_i$ are alternately heavy and light.
The triangle blossom $B_1$ has a slightly different structure
from the others. Changing it to have the same structure gives a similar
construction, with multiple petalevels. We do not present it
because that structure cannot
be grown using the algorithm of
Fig.\ref{BlockingAlgFig}.
Fig.\ref{StartRecPetalFig} shows how the algorithm grows the search structure
of Fig. \ref{RecPetalFig}.

Note also that a similar example uses only light blossoms. So
our construction can be modified for ordinary matching.
The given construction has the advantage of being smaller.

\bigskip

\begin{figure}[t]
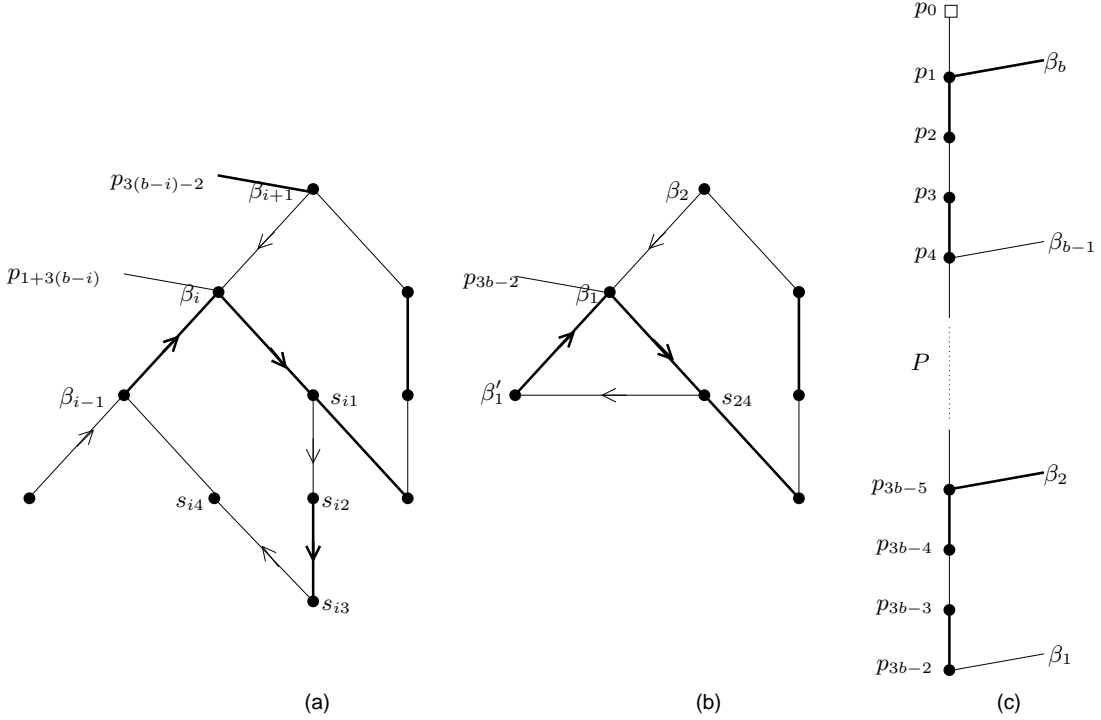

\centering
\input RecPetalDefn.pstex_t
\caption{General structure of $EG(b)$:
(a) Blossoms $B_i$ and $B_{i+1}$, $i>1$ odd:
Blossom $B_{i-1}$ is extended to $B_i$ by a path of 5 edges,
$\beta_{i-1},\beta_i, s_{i1},s_{i2},s_{i3},s_{i4}$.
Arrows give edges on the alternating trail for $l_o(\beta_i)$.
$B_i$ extends to $B_{i+1}$ by a similar path, with similar alternating trail.
For any $j$ (odd or even), $s_{j4}= s_{j-1,1}$ and cross edge
$\beta_j p_{1+3(b-j)}$
is unmatched iff $j$ is odd.
(b) Base case of the construction:
Blossoms $B_1$ and $B_2$
are defined like part (a) except $B_1$ is replaced with a triangle as shown.
(c) $P$ is an alternating path of length $3b-1$.
The subpath  of $P$ from $p_0$ to
$p_{1+3(b-i)}$, followed by the
cross edge to
blossom base $\beta_i$, is also alternating.}
\label{RecPetalDefnFig} 
\end{figure}

Fig.\ref{RecPetalDefnFig} illustrates the general structure of
the example graph $EG$.
In addition we assume $f(p_0)=f(\beta_b)=1$. The remaining
vertices are saturated, as implied by the figure, and their $f$-value is
1 or 2.
The figure implies the following expressions that combine to define
the augmenting trail of
a cross edge $\beta_i p_{1+3(b-i)}$.

Let $j$ be the io-type
$o$ ($i$) for $i$ odd (even). Let $S_i= s_{i1}s_{i2}s_{i3}s_{i4}$.
For every $i\ge 1$,

\begin{equation}
\label{lbetaiEqn}
l_j(\beta_i)= \beta_b\ldots \beta_i S
\beta_1' \beta_1\ldots \beta_i
\end{equation}
\noindent
where
\begin{equation}
\label{SDefnEqn}
S=\begin{cases}
S_iS_{i-1}\ldots  S_2&i>1\\
s_{24}&i=1.
\end{cases}
\end{equation}
This trail illustrated
in Fig.\ref{RecPetalFig} for $i=2$ and \ref{RecPetalDefnFig}(a)
for $i$ odd.

To calculate the length of $\beta_i$'s trail  first note
$|S|=3(b-i)$.  
So
$|l_j(\beta_i)| = (b-i)+1+ 3(i-1) + 1 + 1 +(i-1) = b +3i -1$.

The augmenting trail for $\beta_i$ is
\begin{equation}
\label{AugTrailForBetaiEqn}
l_j(\beta_i) p_{1+3(b-i)}\ldots p_0.
\end{equation}
Its length is
$(b+3i-1) + 1 + (1+3(b-i))= 4b+1$ as desired.

It is clear that the graph has size $\Theta(b)$, and 
\eqref{lbetaiEqn} implies there are $\Theta(b^2)$ petalevels.
It only remains  to prove  the augmenting trails
of \eqref{lbetaiEqn}
are all sats,
i.e., $s=4b+1$. Appendix \ref{ExampleGraphAppendixSec}
shows this by exhibiting the optimum dual values.
Here we present a path-tracing argument for greater intuition.

Let $T$ be a sat. The degree constraints imply
$T$ contains both free vertices $p_0$ and $\beta_b$.
So $T$ contains a cross edge $p_{1+ 3(b-i)}\beta_i$.
The alternation of 
$T$ implies 
similar properties on  both sides of
the cross edge:

\begin{proposition}
\label{BothSidesAltProp}
(a) $T\cap P$ is the path $p_0,\ldots, p_{1+ 3(b-i)}$.

(b) For $i=1$ $T$ contains
$\beta_1 \beta_1' s_{24}\beta_1$ or its reverse.

(c) For $i>1$ if $T$ contains the directed edge $s_{i1}s_{i2}$ then it contains
$S\beta_1' \beta_1\ldots \beta_i$. ($S$ is defined as in \eqref{SDefnEqn}.)
\end{proposition}

\begin{proof}
(a) A cross edge directed to $p_j$
alternates with $p_jp_{j-1}$, but not $p_{j+1}p_j$
(see Fig. \ref{RecPetalDefnFig}(c)).
Using this, a simple downward induction (starting at $j= p_{1+ 3(b-i)}$)
shows $T$ contains directed edge   $p_jp_{j-1}$.

The arguments for (b)--(c) are similar to (a):
Arriving at a vertex $v$ on a directed edge $uv$,
alternation implies there is only one possibility for the next edge.

(b) Examining Fig. \ref{RecPetalDefnFig}(b), we 
use the nonalteration of directed edge $s_{24}s_{23}$.

(c)
Examining Fig. \ref{RecPetalDefnFig}(a)-(b), we successively
use nonalteration of
directed edges $s_{j4}\beta_{j-1}$, $s_{24}\beta_1$,
$\beta_js_{j+1,4}$.
\end{proof}

Part (a) of the lemma shows 
$T$ consists of the trail $p_0\ldots p_{1+ 3(b-i)}\beta_i$
and a trail from $\beta_i$ to $\beta_b$, that begins with
an edge of opposite M-type from the cross edge
$p_{1+ 3(b-i)}\beta_i$.
We claim that trail
must be $l_j(\beta_i)$.
\eqref{AugTrailForBetaiEqn}
then shows $s=4b+1$ as desired.

Before proving the claim we show another simple property
(self-evident from Fig. \ref{RecPetalDefnFig}(a)-(b)).

\begin{proposition}
\label{ShtstPthsiProp}
A shortest path tree for $B_b$ is
$E(B_b)- \set{s_{i3}s_{i4}}{i>1} - s_{24}\beta_1'$.
\end{proposition}

\begin{proof}
First observe that for any $i>1$, the directed edge $s_{i3}s_{i4}$
is not in a shortest path
tree of $B_i$. This follows since the path $\beta_i\beta_{i-1}s_{i4}$
is shorter than $\beta_i s_{i1}s_{i2}s_{i3}s_{i4}$.
Using this, an easy induction shows no edge  $s_{j3}s_{j4}$
is in a shortest path tree for $B_i$. This easily gives the
proposition.
\end{proof}

For $i=1$ a shortest path from $\beta_1$ to $\beta_b$ is
$\beta_1\ldots \beta_b$.
For every $i>1$,
a shortest path from $s_{i1}$ to $\beta_b$ is
$s_{i1}\beta_i\ldots \beta_b$.

\bigskip

\noindent
{\bf Proof of the Claim:}
We first assume $1<i<b$. The two exceptions are minor variants,
which we  treat after the main case.

$T$ enters blossom $B_i$ on cross edge $p_{1+ 3(b-i)}\beta_i$.
To reach $\beta_b$ it leaves $B_i$ either on edge $\beta_i\beta_{i+1}$
or $s_{i+1,4}s_{i+1,3}$.
Observe that in both cases $T$ contains an alternating
trail from $\beta_i$ to the edge  $s_{i1}s_{i2}$:
To leave on $\beta_i\beta_{i+1}$ alteration implies $T$ uses both
edges descending from the cross edge, and alteration at $s_{i1}$ implies
$T$ uses $s_{i1}s_{i2}$ as claimed.
To leave on $s_{i+1,4}s_{i+1,3}$, alteration at 
$s_{i+1,4}=s_{i1}$ also implies $T$ uses $s_{i1}s_{i2}$.

Now Proposition \ref{BothSidesAltProp}(c) shows
$T$ contains 
\[S\beta_1' \beta_1\ldots \beta_i.\]
The rest of $T$ joins
$s_{i1}$ to $\beta_b$. 
Proposition \ref{ShtstPthsiProp}
implies a shortest such path path is
\[s_{i1}\beta_i\ldots \beta_b.\] $T$ can use this path, since it
alternates with $s_{i1}s_{i2}$.
Since $T$ is a sat we can assume $T$ uses it.
Now \eqref{lbetaiEqn} shows the  claim holds.

\bigskip

\noindent Case $i=1$:
After the unmatched cross edge $T$ must contain the 
two matched edges incident to $\beta_1$ (in either order).
So $T$ contains the entire triangle. Then it can follow
the path $\beta_1\ldots \beta_b$. As in the main case we can assume
it is
used. So wlog $T$ 
contains $\beta_1\beta'_1 s_{24} \beta_1\ldots \beta_b$ as desired.

\bigskip

\noindent Case $i=b$:
The matched cross edge implies
$T$ contains an alternating trail beginning and ending
with unmatched edges, i.e., $\beta_b\beta_{b-1}$
and $\beta_b s_{b1}$. The latter implies $T$ contains
$s_{b1}s_{b2}$. The rest of the argument is the same as $i<b$.
\hfill $\spadesuit$

\section{The Simple Graph Bound}
\label{BoundSec}
As shown in the previous section the $n$ vertices of a given graph
may have $\Theta(n^2)$ different petalevels. 
Most of these must be discarded in our desired level graph.

\paragraph*{Motivation}
We begin by discussing an approach to define a small level graph
that is tempting but, to the best of our knowledge, fails.

Like \cite{DHZ, DPS}  it 
is based on classifying each blossom as ``big'' or ``small'',
depending on whether its ``size'' (appropriately defined) is
$\ge$ or $<$ $n^{1/3}$.
Assuming disjointness there are $\le n/n^{1/3}=n^{2/3}$ big blossoms,
so augmenting trails through their base edges can be ignored.
A vertex $v$ in a small blossom $S$ can be traversed on an edge
of $\gamma(S)$ at most to $n^{1/3}$ times, so only $n^{1/3}$
petalevels are needed for $v$. And over $n^{2/3}$ levels, some level
has $\le (n \times n^{1/3})/n^{2/3} = n^{2/3}$ such petalevels, so
$\le n^{2/3}$ augmenting trails pass through that layer of $LG$.
Unfortunately
it does not seem possible to complete this approach. For instance
the edges of $\delta(S)$ cannot be accounted for. In fact our approach
ignores the notion of big blossoms entirely.

Instead, our approach resembles previous definitions of the level graph:
Every vertex is represented at most twice in the level graph (just once
for the previous definitions). An important difference is that our level
graph depends on the collection \cT\ of augmenting trails being analyzed.
In contrast previous definitions have a unique level graph.

To begin consider a graph with edge weights \eqref{wDefnEqn}.
All references to a matching,
blossoms, dual variables, etc. refer to some fixed structured matching.
(The structured matching
will be provided by the algorithm of Fig.\ref{BlockingAlgFig}.)

We start with some motivation. We shall see the following formula
for the  ordinary level of a vertex
$v$ and any $j \in \{\rm{i,o} \}$:
\begin{equation}
\label{LevelDefn}
l_j (v) =
\begin{cases}
y(v)-y(\phi)&j = o\\
1-(y\{v, \phi\}+ z(v)) &j=i
\end{cases}
\end{equation}
where 
\[z(v)=z\set{B}{v\in B}.\]
The formula is valid for any search structure containing the
corrsponding trail $P_j(v)$.

Reading $P_j(v)$ in reverse, note that it enters every blossom containing
$v$ through its base edge (e.g., $P_j(\beta_1)$ for either $j$
in Fig.\ref{RecPetalFig}).
When $v$ occurs on a petalevel in some $P_j(v')$,
a blossom containing $v$ can be entered on a petal
(e.g., $v=\beta_1$ in $P_j(\beta_2)$ in Fig.\ref{RecPetalFig}).
Equation \eqref{LevelDefn} generalizes to give the length of petalevels
as follows.

Consider an alternating trail $T$ from free
vertex $\phi$ to $v$, ending in an edge of io-type
$j$. 
Let $\iota$ be the ``entrance sequence'' of $v$ in $T$,
i.e., for every positive blossom $A$ containing $v$, $\iota_A\in \{b,p\}$
is $b$ if $T$ enters $A$ on its base edge,  $p$ if
it enters on a petal ($\iota$ is mnemonic for ``in'' value).
We write $\iota_A(v)$ if the vertex is not clear.
Define the petalevel
\begin{equation}
\label{PetalevelDefn}
l_j (v,\iota) =
\begin{cases}
y(v)-y(\phi) + z_p(v,\iota)&j = o\\
1-(y\{v, \phi\}+ z_b(v,\iota)) &j=i,
\end{cases}
\end{equation}
where for $c\in \{b,p\}$,

\bigskip

\hskip120pt $z_c(v,\iota)=z\set{A}{v\in A, \iota_A=c}$.

\bigskip

\noindent
When $T$ enters and leaves
$A$ more than once, $\iota_A$ is defined
by the last entry. 
Here the dual values $y,z$ are for any structured matching.
As an example in Fig.\ref{RecPetalFig} vertex $\beta_1$ in $P_j(\beta_2)$
has $l_o(\beta_1,\iota)= -2 -(-8) + 2 = 8$.

It is easy to see that \eqref{PetalevelDefn} becomes 
\eqref{LevelDefn} when $T$ is $P_j(v)$.
Note that an atom $v$ has an empty
$\iota$.
For nonatomic $v$, both of its ordinary levels $l_j(v)$ have
$\iota_A=  b$  for every blossom $A$ containing $v$.
Note also that
any entrance sequence $\iota_\cdot(v)$ has
$l_j(v,\iota)\ge l_j(v)$.

\eqref{LevelDefn} and  \eqref{PetalevelDefn}
can be proved by a ``blossom-tracking'' argument similar to
\cite{G17} for ordinary matching. We omit this in this paper,
since we can always rely on our
more powerful petalevel-tracking approach.
We proceed to explain how we 
track an augmenting trail $T$ (for the level graph).

\paragraph*{Natural petalevels}
We will present two versions of the petalevels of
\eqref{PetalevelDefn}. 
The {\em natural petalevels}
define $\iota$ using the following rule.
The dummy vertex $\varepsilon$ leading to free vertices is not in
any blossom and so has an empty entrance sequence. 
Suppose $T$  advances on an edge $uv$, with $\iota(u)$ already defined.
For any positive blossom $A$ containing $v$,
\begin{equation}
\label{EntranceDefn}
\iota_A(v) =
\begin{cases}
\iota_A(u) & u\in A\\
b & u\notin A, uv =\eta(A)\\
p & u\notin A, uv \ne\eta(A).
\end{cases}
\end{equation}
These natural petalevels are consistent with the description of petalevels in
\eqref{PetalevelDefn}
Note this definition makes  $\iota(v)$  empty if $v$ is atomic.  Also note
that the entry for a base edge is consistent: For blossoms $A'\pcon A$
with the same base edge $uv=\eta(A)=\eta(A')$, $\iota_{A'}(v)
=\iota_{A}(v)$.

We will analyze the natural petalevels before discussing the second version of
petalevels.
The following lemma is the basis of the  crucial
property that $T$ goes from one level of $LG$
to the next, unless
it falls back to a lower level. The lemma gets extended
to the variant petalevels in the corollary below.

We  associate an  i/o type with an edge by using
its M-type and assuming we are advancing along the edge.
Specifically for an M-type $\mu \in \{M, \o M\}$  define
\begin{equation}
 j(\mu) =\begin{cases}
i& \mu=\o M\\
o& \mu= M.
\end{cases}
\end{equation}
When considering an edge $uv$ we usually define $j=j(\mu(uv))$.
Also for $j\in \{i,o\}$ $\bj$ denotes the opposite i/o value from
$j$. Similarly for  $\bj(\mu(uv))$.

\begin{lemma}
\label{PetalevelAdvancementLemma}
Consider an edge $uv$ in an augmenting trail $T$, where all petalevels are natural.

(a) For $j=j(\mu(uv))$,
\begin{equation}
\label{PetalevelAdvancementEqn}
l_j(v,\iota(v))\le l_{\bj}(u,\iota(u)) +1.
\end{equation}

(b) Equality holds in (a) iff $uv$ is tight and it alternates at every
positive blossom that it leaves, i.e., for every positive blossom $A$
with $u\in A \not\ni v$, either $\iota_A(u)=b$ or $uv=\eta(A)$.
\end{lemma}

\begin{proof}
The two possibilities $uv$ matched and unmatched are symmetric, so
we  cover them in parallel.
We begin by reducing the lemma  to symmetric inequalities
\eqref{ZgzzEqn} and 
\eqref{ZgzzEqnU}, and their tightness.

\case{$uv\in M$}
The lemma's inequality is $l_{i}(u,\iota(u))+1 \ge l_o(v,\iota(v))$, i.e.,
$1-\big(y\{u, \phi\} +z_b(u,\iota(u))\big) +1 \ge  y(v)-y(\phi)+
z_p(v,\iota(v))$.
Rearrange this to the equivalent inequality
\begin{equation}
  \label{PetalLevelEquivalenceEqn}
  2\ge y\{u,v\}+z_b(u,\iota(u))+ z_p(v,\iota(v)).
\end{equation}

Since $uv$ is undervalued,
\begin{equation}
\label{2gyzeqn}
  2\ge y\{u,v\} + z(uv).
\end{equation}
We will show
\begin{equation}
  \label{ZgzzEqn}
  z(uv)\ge z_b(u,\iota(u))+ z_p(v,\iota(v)),
\end{equation}
Part (a)  follows from \eqref{2gyzeqn} and \eqref{ZgzzEqn}.
So we need only prove the latter for part (a).
Equality in \eqref{PetalevelAdvancementEqn} is equivalent to equality in
\eqref{2gyzeqn} and \eqref{ZgzzEqn}.
So for (b) we need only prove equality in \eqref{ZgzzEqn} is equivalent
to (b)'s second condition.

\case{$uv\in \o M$}
We wish to prove $l_{o}(u,\iota(u))+1 \ge l_i(v,\iota(v))$, i.e.,
$y(u)-y(\phi)+ z_p(u,\iota(u))+1 \ge 1-\big(y\{v, \phi\} +z_b(v,\iota(v))\big).$
Rearrange this to the equivalent inequality
\begin{equation}
  \label{PetalLevelEquivalenceEqnU}
  y\{u,v\}+z_p(u,\iota(u))+ z_b(v,\iota(v))\ge 0.
 \end{equation}

Since $uv$ is dominated,
\begin{equation}
\label{2gyzeqnU}
y\{u,v\} + z(uv)\ge 0.
\end{equation}
We will show
\begin{equation}
  \label{ZgzzEqnU}
z_p(u,\iota(u))+ z_b(v,\iota(v))\ge z(uv).
\end{equation}
\eqref{PetalevelAdvancementEqn}
follows from \eqref{2gyzeqnU}--\eqref{ZgzzEqnU}, so we need only prove
\eqref{ZgzzEqnU} for (a). Similarly 
equality in \eqref{PetalevelAdvancementEqn} amounts to equality in
the two above inequalities.
So (b) amounts to showing
equality in \eqref{ZgzzEqnU} is equivalent
to (b)'s second condition.

\bigskip

We will show that the desired inequalities
\eqref{ZgzzEqn}
and \eqref{ZgzzEqnU}
hold blossom-by-blossom, i.e.,
any blossom $A$ contributes at least as much to
the LHS as the RHS.
To do this, in the context of the chosen blossom $A$ we let
$LHS$ be $A$'s contribution to the LHS of \eqref{ZgzzEqn}
(respectively \eqref{ZgzzEqnU})
and similarly
for $RHS$, when discussing the set $M$ (resectively $\o M$).
We will show $LHS\ge RHS$.
In addition we will show some $A$ has
$LHS>RHS$ iff $uv$ does not alternate at $A$, more precisely,
$u\in A \not \ni v$ with $\iota_A(u)=p$ and $uv$ a petal of $A$.
This clearly  implies part (b) of the lemma.

\bigskip

A blossom $A$ contributing to a term in \eqref{ZgzzEqn} or \eqref{ZgzzEqnU}
contains at least one of $u,v$.
(In particular we are done if neither vertex is in a blossom.
A vertex not in a blossom is either atomic or it does not belong to \S..)
We consider three cases,
$uv\in \gamma(A)$, $uv=\eta(A)$, or $uv$ a petal of $A$.

\case{$u,v\in A$}
$A$ contributes $z(A)$ to $z(uv)$. We show it makes the same contribution
to the other side of \eqref{ZgzzEqn} and \eqref{ZgzzEqnU}.

\eqref{EntranceDefn} for this case implies $\iota_A(u)=\iota_A(v)$.
This value $\iota_A(u)$ 
is either b or p. Thus  $z(A)$
contributes to exactly one of the terms $z_b(u,\iota(u))$,
$z_p(v,\iota(v))$ in \eqref{ZgzzEqn},
and exactly one of the terms $z_p(u,\iota(u))$,
$z_b(v,\iota(v))$ in \eqref{ZgzzEqnU}.

So $LHS=RHS=z(A)$ as desired.

\case{$uv=\eta(A)$}
First assume $uv$ is matched. We will show $LHS=RHS=0$.
The case makes $uv\not\in I(A)$, so $LHS=0$.

Suppose $v=\beta(A)$.
Thus $T$ enters $A$ on its base $uv$.
So $T$ does not enter $A$ on a petal, i.e.,
it does not contribute to the RHS term $z_p(v,\iota(v))$.
Also $u\notin A$, so $A$ does not contribute to
the RHS term $z_p(u,\iota(u))$. So $RHS=0$ as desired.

Suppose $u=\beta(A)$.
Thus $T$ leaves $A$ on its base $uv$.
So $T$ enters $A$ on a petal.
Thus $A$ does not contribute to the RHS term
$z_b(u,\iota(u))$.
Also $v\notin A$, so $A$ does not contribute to
the RHS term $z_p(v,\iota(v))$.
So $RHS=0$ as desired.

Next assume $uv$ is unmatched.  Opposite from before we show $LHS =
RHS =z(A)$.  The case makes $uv\in I(A)$,
so $RHS=z(A)$.

Suppose $v=\beta(A)$.  Thus $T$ enters $A$ on its base $uv$.  So $A$
contributes to the LHS term $z_b(v,\iota(v))$.  Also $u\notin A$, so
$A$ does not contribute to the LHS term $z_p(u,\iota(u))$. So
$LHS=z(A)+0=z(A)$ as desired.

Suppose $u=\beta(A)$.  Thus $T$ leaves $A$ on its base $uv$.  So $T$
enters $A$ on a petal.  This makes $A$ contribute to the
LHS term $z_p(u,\iota(u))$.  Also $v\notin A$, so $A$ does not
contribute to the LHS term $z_b(v,\iota(v))$.  So $LHS=z(A)+0=z(A)$ as
desired.

\case{$uv$ a petal of $A$}
First assume $uv$ is matched.  Clearly $uv\in I(A)$ so $LHS =z(A)$.
Since $A$ contains only one of the vertices $u,v$, only one of the RHS
terms $z_b(u,\iota(u))$, $z_p(v,\iota(v))$ can be positive.  Thus
$LHS \ge RHS$ as desired.

Suppose $u\in A$.  Note the RHS term $z_b(u,\iota(u))$ has a
contribution from $A$ iff $\iota_A(u)=b$, which may or may not hold
for $T$.
$\iota_A(u)=b$ implies $RHS=z(A)=LHS$.
Also $T$ alternates at $A$, as desired for (b).
$\iota_A(u)=p$ implies $RHS=0<z(A)=LHS$. $T$ does not
alternate at $A$. Again (b) holds.

Suppose $v\in A$.   $A$
contributes to the RHS term $z_p(v,\iota(v))$, so $LHS=RHS$.

Next assume $uv$ is unmatched. Clearly $uv\not\in I(A)$ so $RHS =0$.
Thus $LHS \ge RHS$ as desired.

Suppose $u\in A$.  The LHS term $z_p(u,\iota(u))$ has a
contribution from $A$ iff $\iota_A(u)=p$.
$\iota_A(u)=b$ implies $RHS=0=LHS$.
$T$ alternates at $A$, as desired for (b).
$\iota_A(u)=p$ implies $LHS=z(A)>0=RHS$. $T$ does not
alternate at $A$. Again (b) holds.

Suppose $v\in A$.   $A$ does not
contributes to the RHS term $z_b(v,\iota(v))$, so $LHS=RHS=0$.
\end{proof}

\paragraph*{Shortened petalevels}
As noted there are too many natural petalevels
to use in the level graph.
But some of them are extraneous.
The following property
is the source of the  redundancy.

\begin{proposition}
\label{ShortenDefProp}
Let $\iota$ and $\iota'$ be in-values for vertex $v$
that are identical except that $\Delta$ positive blossoms $A$
have $\iota_A=p$ and $\iota'_A=b$. Then for any $j\in \{i,o\}$,
$l_j(v,\iota')\le l_j(v,\iota) -2\Delta$.
\end{proposition}

\begin{proof}
If $v$ is outer $A$ contributes $z(A)$ to the $\iota$ petalevel
and 0 to the $\iota'$ version. If $v$ is inner the contribution is
0 to the  $\iota$ petalevel
and $-z(A)$  to the $\iota'$ version.
\end{proof}

For clarity we  write $LG^*$ to denote our version of
the level graph (using $LG$ to denote generic level graphs).
The petalevels used in $LG^*$
sometimes make the change of the proposition, i.e., p entries are
changed to $b$. In keeping with the proposition we call this change
a {\em shortening}.
Intuitively, shortening destroys progress of $T$ towards the goal,
the last level of $LG$. But that progress is not needed, since $T$ advances
from the shortened petalevel to the goal only one level at a time.

We integrate shortening into the tracking as follows.  To set the
stage, we are given a collection \cT\ of edge disjoint augmenting
trails.  Start by choosing an arbitrary orientation for every
\cT-trail.  For each positive blossom $A$ we will define a unique
value $\iota_A$ such that every vertex $v\in A$ has
$\iota_A(v)=\iota_A$ in each of its petalevels. Specifically we define
$\iota_A=b$ if some trail $T\in \cT$ enters $A$ on its base.
Otherwise $\iota_A=p$.

Observe this is well-defined, since any base edge $\eta(A)$ is on at
most one \cT-trail.  The effect of this rule is to change an in-value
$p$ to $b$ when $T$ enters $A$ on a petal, but some other \cT-trail
enters $A$ on its base.

This translates to the new version of 
\eqref{EntranceDefn}: 
\begin{equation}
\label{ShortenedEntranceDefn}
\iota_A(v) =
\begin{cases}
\iota_A(u) & u\in A\\
b & u\notin A, \text{ some \cT-trail enters $A$ on $\eta(A)$}\\
p & u\notin A, \text{ no \cT-trail enters $A$ on $\eta(A)$.}
\end{cases}
\end{equation}
This change means that in-values no longer correspond exactly to
the direction a blossom is entered. The reader should bear this distinction
in mind throughout the rest of this section. Now we show the lemma for
natural petalevels is preserved by shortening.

\begin{corollary}
\label{PetalevelAdvancementCor}
Consider a graph where 
all petalevels are defined using rule \eqref{ShortenedEntranceDefn}.
Inequality \eqref{PetalevelAdvancementEqn}
continues to hold.
\end{corollary}

\remark{The shortening affects the shortened petalevel
as well as 
all of the following petalevels. Because of the latter
the corollary
does not immediately follow from
Proposition \ref{ShortenDefProp}.}

\begin{proof}
As in the lemma consider an edge $uv$.
We will refer to the proof of the lemma, explicitly pointing out any changes
needed for the corollary. All case references below are to the corresponding case in the lemma's proof.

In the first two cases {$uv\in M$}, {$uv\in \o M$},
the reductions to
 \eqref{ZgzzEqn} and \eqref{ZgzzEqnU} are unchanged.
 The $z_p$ and $z_b$ quantities are now interpreted using 
\eqref{ShortenedEntranceDefn} for $\iota$-values.

\case{$u,v\in A$}
The proof is unchanged:
The claim that $\iota_A(u)=\iota_A(v)$
follows from the first case of \eqref{EntranceDefn},
which is also the first case of \eqref{ShortenedEntranceDefn}.

\case{$uv=\eta(A)$}
First assume $uv$ is matched.

Suppose $v=\beta(A)$. 
Thus $T$ enters $A$ on its base $uv$.
The second case of \eqref{ShortenedEntranceDefn} shows
$\iota_A(v)=b$. Thus
$A$
does not contribute to the RHS term $z_p(v,\iota(v))$, as in the lemma.

Suppose $u=\beta(A)$.
As in the lemma
$T$ enters $A$ on a petal, say $rs$. 
As in the lemma
the $RHS$ term $z_b(u,\iota(u))$ is 0, but now the
following argument is used.
The first case of \eqref{ShortenedEntranceDefn} implies
$\iota_A(u)=\iota_A(s)$. 
No \cT-trail enters $A$ on its base
(since $T$ leaves $A$ on its  base
edge $\eta(A)=uv$, and \cT\ uses edge $uv$ only once). Thus 
\eqref{ShortenedEntranceDefn} defines $\iota_A(s)$ as $p$. So
 $\iota_A(u)=p$ and
$z_b(u,\iota(u))=0$ as desired.

Now assume $uv$ is unmatched.

Suppose $v=\beta(A)$.  As in the lemma this means
$T$ enters $A$ on its base $uv$.  $T$ satisfies the second line of
 \eqref{ShortenedEntranceDefn}, i.e., $u\notin A$ and $T$ enters $A$ on its base.
So $A$
contributes to the LHS term $z_b(v,\iota(v))$.  As in the lemma
$A$ does not contribute to the LHS term $z_p(u,\iota(u))$.

Suppose $u=\beta(A)$.  As in the lemma this means
$T$ leaves $A$ on its base $uv$. Let $rs$ be the edge of $T$
that enters $A$ before leaving on $uv$. \cT\ uses $uv$ only once, so
$\iota_A(s)=p$ by the third line of \eqref{ShortenedEntranceDefn}.
The first line implies $\iota_A(u)=p$. Thus
$A$ contributes to the
LHS term $z_p(u,\iota(u))$, as desired.

\case{$uv$ a petal of $A$}
First suppose $uv$ is matched. 
If $u\in A$ the argument does not change:
The RHS term $z_b(u,\iota(u))$ has a
contribution from $A$ iff $\iota_A(u)=b$, which may or may not hold
for $T$. If it does then 
$RHS=z(A)=LHS$.

If $v\in A$ then $T$ enters $A$ on a petal.
The second or third line
of \eqref{ShortenedEntranceDefn} applies. The third line keeps
$\iota_A(v)=p$ so as in the lemma,
$A$ contributes to the RHS term $z_p(v,\iota(v))$.
The second line means $A$ does not contribute to the RHS and
$LHS >RHS$.

Next assume $uv$ is unmatched. As in the lemma
this makes $RHS =0$ so
$LHS \ge RHS$ as desired.
In more detail suppose $u\in A$.
$T$ may or may not have $\iota_A(u)=p$. The former gives strict inequality
$LHS > RHS$. Suppose $v\in A$. $T$ may or may not have $\iota_A(v)=b$.
The former gives strict inequality.
\end{proof}

\paragraph*{Analysis}
We prove properties of shortened petalevels that  lead to the definition
of our level graph and its correctness.

The first fact will be used to define the last level of our level graph.
Consider an iteration of the matching algorithm Fig.\ref{BlockingAlgFig}.
The algorithm uses the $f$-matching search algorithm
of \cite{G18} to
find an augmenting trail.
The search ends with a structured matching for
an augmenting trail.
Let $y(\phi)$ denote the $y$-value of free vertices in
this structured matching.
Define 
$L=-y(\phi)$.

\begin{lemma}
\label{EndAtLastLevelLemma}
The last vertex of any  \cT-trail has petalevel $1+2L$.
\end{lemma}

\remark{The lemma does not preclude multiple visits to this petalevel.}

\begin{proof}
Suppose $T\in \cT$ starts at a free vertex $\phi$
and  ends at a free vertex $\phi'$ on an odd petalevel
$l_i(\phi',\iota(\phi'))$.
The edge leading to $\phi'$
is unmatched, and possibly in various blossoms $A$ with base vertex $\phi'$.
(It is also possible that $\phi=\phi'$.)
The definition of petalevels
\eqref{PetalevelDefn} gives
$l_i(\phi',\iota(\phi'))=
1-(y\{\phi', \phi\}+ z_b(\phi',\iota(\phi')))$.
We will show the $z$ term equals 0. This makes the petalevel equal to
$1-y\{\phi', \phi\}=1+2L$, as claimed.

To analyze the $z$ term we start with this observation: Let $\phi$ be
the base vertex of a positive blossom $A$. (There may be many such
$A$.) $\phi$ is on at most one \cT-trail $T$, and $\phi$ is the first
or last vertex of $T$ but not both. In proof, since $A$ is positive
the $\phi\phi$-trail forming $A$ is not augmenting. Thus
$def(\phi)=1$. Thus $\phi$ occurs at most once as an end of
a \cT-trail.

Now return to the given trail $T$ ending at $\phi'$.  Take any
positive blossom $A$ with base vertex $\phi'$. $T$ ends on the base
edge of $A$, $\eta(A)=\varepsilon \phi'$. The observation implies
no \cT-trail enters $A$ on its base $\eta(A)$.
Thus \eqref{ShortenedEntranceDefn} shows $\iota(A)=p$. Hence
$z_b(\phi',\iota(\phi'))=0$ as desired.
\end{proof}

Recall the lemma tracks \cT-trails with shortened petalevels.
A version of the lemma with natural petalevels is useful:

\begin{corollary}
\label{EndAtLastLevelCor}
Tracking an augmenting trail with
natural petalevels, its last vertex has petalevel $1+2L$.
\end{corollary}

\begin{proof}
The argument is simpler -- it is obvious that $T$ enters
a blossom containing $\phi'$ on a petal.
\end{proof}

We now present the details of our level graph.
We call it
$LG^*$, to distinguish it from references to
generic level graphs $LG$.
We use the term {\em node} to refer to vertices in the level graph.
Define the last level of $LG^*$ to be $1+2L$.

Each edge of $T$ advances at most one level in $LG^*$. We conclude every
$\cT$-trail uses an edge in every layer of $LG^*$.  (N.B. The edges of $T$
that advance from level 0 to $2L+1$ need not form an
augmenting trail or even contain one: Loose edges retreat in $LG$ and lead to duplicate advances. $T$ may leave the  search graph \S. and reenter at a different point, etc.)

We complete the definition of $LG^*$ by specifying the
petalevels that it uses. 
Every vertex $v$ has at most
two petalevels in $LG^*$, $l_j(v,\iota(v))$.
Here $j$ ranges over the io-types
i,o, and the in-values $\iota(v)$ are consistent
with our shortening rule: For every positive blossom $A$ containing $v$,
\begin{equation}
\label{BigBlossomLGDefn}
\iota_A(v)=
\begin{cases}
b&\text{some \cT-trail enters $A$ on its base $\eta(A)$}\\
p&\text{no \cT-trail enters $A$ on its base $\eta(A)$}.
\end{cases}
\end{equation}
If $v$ is atomic, i.e., not in any positive blossom, these $\le 2$ petalevels
are ordinary levels $l_j(v)$, by \eqref{LevelDefn}.
$LG^*$ contains every possible edge that joins two such nodes
on consecutive levels.

We show this level graph $LG^*$ tracks every \cT-trail $T$
from level 0 to level $2L+1$.
To be precise assume 
 \eqref{ShortenedEntranceDefn} is used to define the petalevels
 that track $T$.
Let $L(T)$ denote the set of edges of $T$ that are in the final
set of tracked advances.
($L(T)$ need not include 
every forward edge of $T$. For example a petal $uv$ that enters a positive
blossom becomes a backup edge if it gets shortened.)

\def\myeqnumLGT(#1)#2{\bigskip\noindent{$(#1)$\hskip20pt $#2$}\bigskip}

\begin{lemma}
Suppose edge $uv\in L(T)$ advances 
from level $l_{\bar j}(u,\iota(u))$ to $l_j(v,\iota(v))$
(where these petalevels are defined using \eqref{ShortenedEntranceDefn}).
The level graph $LG^*$ 
has a corresponding edge from node $u$ on level
$l_{\bar j}(u,\iota(u))$ to node $v$ on level $l_j(v,\iota(v))$.
\end{lemma}

\begin{proof}
The main argument is to 
show the vertex $v$ has a corresponding
node on level  $l_j(v,\iota(v))$. (Here $v$ is an arbitrary vertex,
so this suffices to show $u$ also has its corresponding vertex.)

First suppose $v$ is an atom,
i.e., not
in any positive blossom.
As previously noted, \eqref{ShortenedEntranceDefn}
and \eqref{LevelDefn} define $l_j(v,\iota(v))$ as $l_j(v)$, and $LG^*$
has been defined to contain the corresponding node.

Next suppose $v$ is contained in various positive blossoms, and fix such a blossom  
$A$.
Suppose $T$ enters $A$ on edge $uv_1$ and follows the trail to $v$,
$v_1v_2\ldots v_k=v$, that is contained
in $A$. \eqref{ShortenedEntranceDefn} defines
 $\iota_A(v_1)$ according to line 2 or 3, i.e., for $x=v_1$,

\bigskip

\noindent $(\dagger)$ \hskip100pt
$\iota_A(x)=b$ iff some \cT-trail enters $A$ on $\eta(A)$.

\bigskip

\noindent
The first line of \eqref{ShortenedEntranceDefn} defines
$\iota_A(v_i)$ as $\iota_A(v_{i-1})$. Thus $(\dagger)$ holds for every $x=v_i$,
including $x=v_k=v$. $(\dagger)$ for $v$
is identical to the definition of  $\iota_A(v)$ in 
\eqref{BigBlossomLGDefn}, as desired.

We conclude that $LG^*$ contains the desired nodes for $u$ and
$v$. By definition
$LG^*$ contains every possible edge in the layer for $uv$,
so it contains the desired edge for $uv$.
\end{proof}

Let $s$ denote the length of a sat in $G$. We will show $s$ is
the number of levels in
$LG^*$, $s=1+2L$. The reason is that the $f$-matching search algorithm
finds a sat. This can be seen from first principles,
but we prefer to use
our tracking of natural petalevels.

Let $T$ be the augmenting trail found by the $f$-matching algorithm in
Fig.\ref{BlockingAlgFig}. Recall that
$T$ is discovered as an alternating path
in the algorithm's search structure \oS..
The subtrail of $T$ through any contracted blossom $B$ of \oS.
is the appropriate trail
$P_i(v,\beta(B))$
(recall Section \ref{BlossomReviewSec}).
The construction of these trails 
(Lemma 4.4 of \cite{G18}) easily shows that
they alternate at every blossom they traverse.
(This includes all subblossoms of $B$.)
We conclude that every edge of $T$ satisfies the conditions of
Lemma \ref{PetalevelAdvancementLemma} (b).
Thus every edge of $T$ advances to the next highest petalevel.

Corollary \ref{EndAtLastLevelCor} shows
$T$ ends at petalevel $1+2L$.
$T$ never visits a petalevel twice,
since it always advances petalevels.
Thus $T$ ends the first time it reaches $1+2L$,
$|T|=1+2L$.
Lemma \ref{EndAtLastLevelLemma}
shows every augmenting trail has length $\ge 1+2L$.
Thus $T$ is a sat
and $s=1+2L$, as desired.

We finally note $LG^*$ has the desired size and our overall efficiency bound
holds:
The $n$ vertices of $G$ have $2n$ corresponding nodes in $LG^*$.
$LG^*$ has $s$ layers.
Some layer $l^*$ of $LG^*$ has $\le (2\times 2n)/s= 4n/s$ nodes.
Thus  $l^*$ has $\le 4(n/s)^2$ edges.
Every \cT-trail uses an edge in $l^*$.
So \cT\ has at most $4(n/s)^2$ trails.
This is the desired key property $(*)$.

\section{The Cardinality Matching Algorithm}
\label{AlgSec}

This section presents the details of our cardinality matching algorithm,
Fig. \ref{BlockingAlgFig}.

\paragraph{Details and correctness}
We use the algorithm of \cite{G18} for the $f$-matching search.
The previous section shows it finds a sat.
It remains only to give the details for finding the
blocking set.

We use the blocking flow algorithm of \cite{G21}.
By definition, that algorithm finds a maximal collection of
disjoint augmenting trails. So a priori it is not guaranteed
to find the \B.-set that we seek. We will prove that in our context,
\cite{G21} does find sats. Thus Fig.\ref{BlockingAlgFig} is correct.

We call \cite{G21} with the set of tight edges and the blossoms
found in the $f$-matching search. (In general \cite{G21} is
called with
a graph and a matching, and an optional set of blossoms.)
The algorithm finds a maximal set of edge-disjoint augmenting trails.
We will show that in our context a trail is augmenting
iff it is a sat. This establishes the validity of our use
of \cite{G21}.

Consider an augmenting trail $T$ returned by \cite{G21}.
 Track the edges of $T$ using natural petalevels.
Lemma \ref{PetalevelAdvancementLemma}(b) shows each edge 
increases the petalevel by 1. Corollary \ref{EndAtLastLevelCor}
shows 
the last vertex of $T$ has petalevel $1+2L=s$.
Thus $T$ is a sat.

\paragraph{Time bound}
Each iteration of our algorithm uses time $O(m)$.
Specifically the time for an $f$-matching search is $O(m)$.
This follows from the algorithm description in \cite{G18},
using data structures given in \cite[Section 5]{G17} for ordinary matching.
(Adapting them for $f$-matching is straightforward. We use
the fact that $f(V)\ge s= 1-2y(\phi)$.)

The time to find the blocking set \B. is $O(m)$ \cite{G21}.
(Note that the trails returned by \cite{G21}
have all positive blossoms contracted. Our algorithm expands
these blossoms --
the trail through a blossom  is the appropriate
$P_i$ trail. These trails can be computed in
linear time \cite [Appendix C]{G21}.)

\begin{theorem}
A maximum cardinality $f$-matching can be found in time
  $O(n^{2/3}\; m)$ for simple graphs
and
$O(\min \{\sqrt {f(V)}, n\}m)$ for a multigraphs. \hfill$\Box$
\end{theorem}

\setcounter{section}{0}
\renewcommand{\thesection}{\Alph{section}}
\renewcommand{\thetheorem}{\Alph{section}.\arabic{theorem}}

\setcounter{equation}{0}
\renewcommand{\theequation}{\Alph{section}.\arabic{equation}}
\section{The Search of the Example Graph}
\label{ExampleGraphAppendixSec}
This section gives shows the $f$-matching algorithm
can grow the
search structure of our example graph.

\paragraph*{The search in \boldmath{$EG(4)$}}
\begin{figure}[t]
\centering
\input StartRecPetal.pstex_t
\caption{Search steps forming blossoms. (a)--(d):
Search structure \S. with $y,z$ values, for consecutive steps
$y(\phi)= -4,-5,-6,-7$. Dashed edges are strictly dominated or strictly underrated and not in \S..
Vertices on the same level
are connected by dotted lines. (e)--(f): Cross edges $e_i$ do not enter
\S., see text.
}
\label{StartRecPetalFig}
\end{figure}

Fig.\ref{StartRecPetalFig} shows how the $f$-matching algorithm grows the
search structure of Fig.\ref{RecPetalFig}.
Parts (a)--(d) show how the blossoms are formed. In addition we
must check that no cross edge enters \S. until the
sat's are discovered (i.e., $y(\phi)=-8$).
This follows from the fact that cross edges are loose (with one minor exception)
before that point. In proof
$\beta_i$ enters \S. strictly before $p_i$.
The next dual adjustment makes an unmatched cross edge
strictly dominated ($y(\beta_i)$ increases by 1, see part (e))
and a matched cross edge strictly underrated
($y(\beta_i)$ decreases by 1, see part (f)).
As long as $\beta_i$ is atomic, dual adjustments keep
the cross edge loose: For an unmatched (matched) cross edge,
$y(p_i)$ decreases by 0 or 1 (increases by 0 or 1), respectively.
After  $B_i$ is formed, dual adjustments 
reduce the slack in $e_i$'s inequality:
If $e_i$ is unmatched
$\H{yz}(e_i)$ decreases by 1 ($0 -1=-1$)
or 2 ($-1 +(-1)=-2$).
If  $e_i$ is matched $\H{yz}(e_i)$ increases by 1 ($0 -1+2=1$)
or 2 ($1 -1+2)=2$). Each
$e_i$ is tight when the sat's are discovered
(recall Fig.\ref{RecPetalFig}). So $e_i$ was strictly dominated
before that point.

\begin{figure}[t]
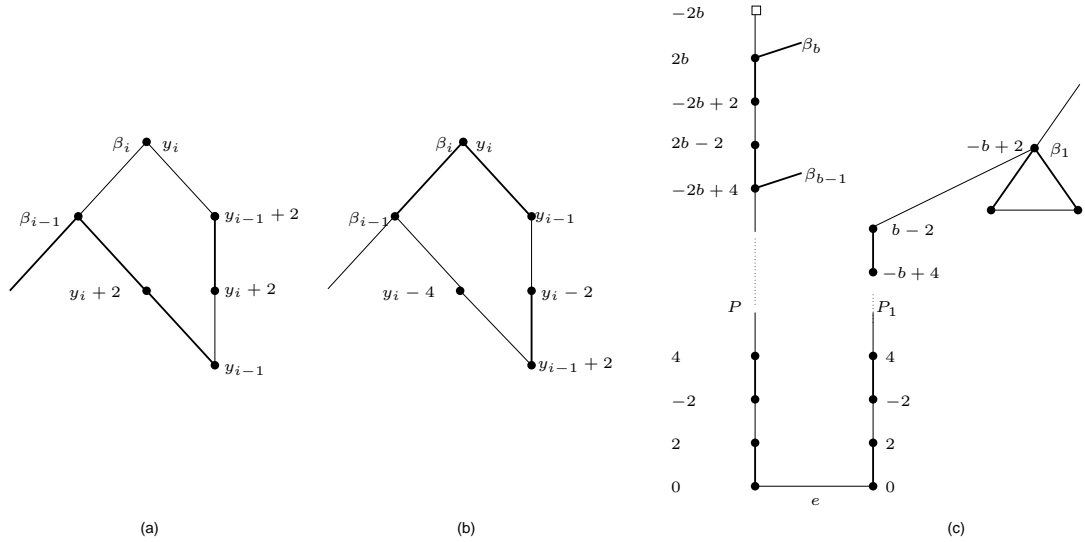

\centering
\input GeneralRecPetal.pstex_t
\caption{Search structure for $EG$.
(a)--(b): $y$-values for the blossoms, for 
 $i$ even (a) and $i>1$ odd (b).
Base vertex $\beta_i$ has $y$-value
$y_i$. 
(c) Edge $e$ joins \oS. paths $P$ and $P_1$, completing
the augmenting trails.
}
\label{GeneralRecPetalFig}
\end{figure}

\begin{figure}[h]
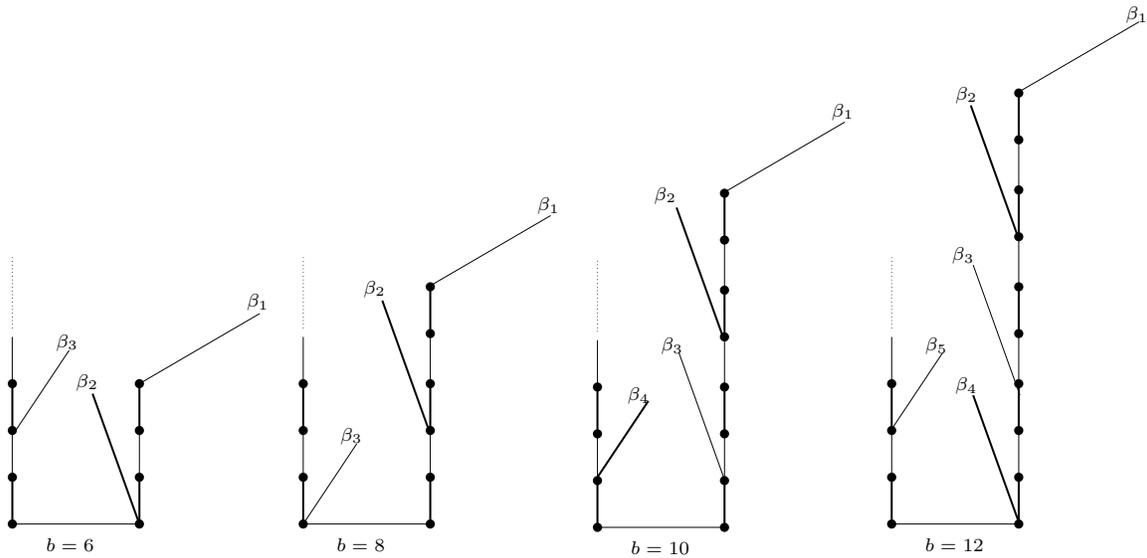

\centering
\input GeneralCrossers.pstex_t
\caption{Cross edge structure in \S..
The three patterns for cross edges switching sides
of paths joined by $e$. $b=12$ has the same pattern
as $b=6$.}
\label{GeneralCrossersFig}
\end{figure}

\paragraph{The general search structure \S.}
Fig. \ref{GeneralRecPetalFig} and \ref{GeneralCrossersFig}
illustrate the final search structure constructed by the
$f$-matching algorithm.
Start by considering the blossom structure,
illustrated in Fig. \ref{GeneralRecPetalFig}(a)--(b).
In the definition of $\H{yz}(e)$ (\eqref{fHyzEqn})
abbreviate the $z$ term to $z(e)$.
The edge $e=\beta_{i-1}\beta_i$ is tight, i.e.,
$y_{i-1}+y_i+z(e)=w(e)$.
This relation, and the tightness of all 6 edges shown
in  Fig. \ref{GeneralRecPetalFig}(a)--(b), implies
all the $y$ values shown.
For example, in (a)
 edge $f=\beta_{i-1}s_{i4}$ has $2 = y_{i-1} + (y_i +2) +z(f)$
(we use  $z(f)=z(e)$, $B_{i-1}$ contributes to both $z$'s).

The relations between the $y_i$'s gives the following constraints.
Let $\beta_0$ and $\beta'_0=s_{11}$ be the two children of $\beta_1$,
and let $y_0$ be their $y$-value.
\begin{eqnarray}
y_i +2 &= &y_{i-2}\text {\hskip 38pt $i\ge 2$ even} \label{iEvenYiEqn}\\
 y_i-4&=&y_{i-2}+2 \text{\hskip 20pt $i\ge 3$ odd} \label{iOddYiEqn}\\
y_0&=&-b\label{0YiEqn}\\
y_1&=&-b+2.\label{1YiEqn}
\end{eqnarray}
In proof the relation $s_{i4}= s_{i-1,1}$ gives \eqref{iEvenYiEqn} and
\eqref{iOddYiEqn}.
The last two lines come from tightness of the triangle blossom:
Tightness of the unmatched edge $\beta_0\beta'_0$
gives $y_0+y_0+2b=0$, i.e., \eqref{0YiEqn}.
Tightness of the matched edge $\beta_0\beta_1$
gives  $y_0+y_1+2b=2$,
so $y_1= 2-2b-y_0$
giving
\eqref{1YiEqn}.
There are no other constraints on the $y_i$'s.

It is easy to see the above system gives these values:
\begin{equation}
\label{iGeneralYiEqn}
y_i=\begin{cases}
-b-i&i \text{ even}\\
-b-1+3i&i \text{ odd}.
\end{cases}
\end{equation}

Using $y_1=-b+2$ verifies the $y$-values shown in
Fig.\ref{GeneralRecPetalFig}(c).
Grow steps in the $f$-matching algorithm give
the two alternating paths that end at $e$, $P$ and $P_1$.
We define $P_1$ to start with the cross edge from $\beta_1$
(ending at $p_{3b-2}$, the inner vertex with $y$-value $b-2$).
Recalling Fig.\ref{RecPetalDefnFig}(c),
we must check that the various cross edges
 $\beta_ip_{1+3(b-i)}$ do not preempt the grow steps for $P_1$ (e.g.,
Fig. \ref{GeneralCrossersFig} indicates potential grow steps for $\beta_2$'s cross edge).
An easy alternative is to enlarge the example graph
with alternating paths $P_i$ similar to $P_1$ but starting at each
base $\beta_i$. The disadvantage is that the example graph
grows to size $\Theta(b^2)$.

\subparagraph{No preemptions}
The grow step to 
vertex $p_{1+3(b-i)}$
occurs either from vertex $\beta_i$ or from a vertex on
$\beta_1$'s grow path. We show that both of these grow steps
become possible when
\begin{equation}
\label{yPhiGoalEqn}
y(\phi)=-b-3i + (i+1 \mod 2).
\end{equation}
Thus the algorithm can choose to grow from $\beta_1$, as desired.

We start with the $\beta_i$ case.
The $b$ blossoms $B_i$ are formed in consecutive dual adjustments.
$B_b$ is formed in the penultimate search, i.e., $y(\phi)=-2b+1$.
So in general
each blossom
$B_i$ is formed when
\begin{equation}
\label{yPhiStartEqn}
y(\phi)= (-2b +1)+ (b-i)= -b-i+1.
\end{equation}
Using \ref{iGeneralYiEqn}, the $y$-value of $\beta_i$ when $B_i$ is formed is
\begin{equation}
\label{iBiFormsYiEqn}
y_i=\begin{cases}
(-b-i)+(b-i+1)= -2i+1&i \text{ even}\\
(-b-1+3i)+(b-i+1)=2i&i \text{ odd}.
\end{cases}
\end{equation}
At that point, assuming $\beta_i$'s cross edge $f=\beta_ip_{1+3(b-i)}$
is still not in \S., $\H{yz}(f)=1+y_i$.

Suppose $i$ is odd. $f$ is unmatched. $f$ is strictly dominated (since
$y_i$ is positive)
and the next $1+y_i$ dual adjustments make $f$ tight,
$\H{yz}(f)=0$. Using \eqref{yPhiStartEqn} these duals adjustments
decrease $y(\phi$ to
$ (-b-i+1)-(1+y_i)=(-b-i+1)-(2i+1)=-b-3i$. Thus
\eqref{yPhiGoalEqn} holds in the odd case.

Suppose $i$ is even. $f$ is matched. $f$ is strictly underated (since
$y_i$ is negative). Since $f$ belongs to $I(B_i)$ and no other $I$ sets,
each dual adjustment increases the value of $\H{yz}(f)$ by 1.
So
the next $2-(1+y_i)$ dual adjustments make $f$ tight.
Using \eqref{yPhiStartEqn} these duals adjustments
decrease $y(\phi$ to
$ (-b-i+1)-(1-y_i)=(-b-i+1)-(2i)=-b-3i+1$. Thus
\eqref{yPhiGoalEqn} holds for $i$ even.

We turn to 
$\beta_1$'s path. We will show that the algorithm
can grow the path $P_1$ shown in Fig. \ref{GeneralRecPetalFig}(c).
View the graph as in
Fig.\ref{RecPetalDefnFig}(c), i.e., $p_j$ is the vertex on $P$ at distance
$j$ from free vertex $p_0$. Execute the search algorithm, always choosing
to 
grow path $P_1$ whenever possible.

Consider the cross edge for $\beta_1$ ending at
$p_{3b-2}$. \eqref{yPhiGoalEqn} shows it becomes tight
when $y(\phi)=-b-3$. This is before any other cross edge becomes
tight, so it is added to \S.. Inductively assume
path $P_1$ has been extended, with  no cross edge from a $\beta_i$, $i>1$
grown,
when $y(\phi)$ gets decreased to the value of \eqref{yPhiGoalEqn}
$-b-3i + (i+1 \mod 2)$, when cross edge $f=\beta_ip_{1+3(b-i)}$ becomes tight.
Note that path $P_1$
extends by 2 edges every other dual adjustment.
More precisely
when
\[y(\phi)=-b-1-2\delta\]
the grow path can be extended
to length $2\delta$, i.e., the $p_{cdot}$ vertices at distance
$2\delta-1$ and $2\delta$ can be added.
(For example the case $\delta=1$ correponds to the above grow step
when $y(\phi)=-b-3$.)
Eyeballing Fig.\ref{RecPetalDefnFig}(c),
the distance of $f$'s end  $p_{1+3(b-i)}$ from the start of $P_1$ is
\begin{equation}
\label{Distance_fEndEqn}
(3b-1)- (1+3(b-i))= 3i-2.
\end{equation}

Suppose $i$ is odd.
Since the distance \eqref{Distance_fEndEqn} is odd,
$P1$ can be extended to this $f$'s end when $2\delta-1= 3i-2$,
$2\delta= 3i-1$.
This corresponds to $y(\phi)=-b-1-2\delta=-b-1-(3i-1)= -b-3i$.
\eqref{yPhiGoalEqn} gives the same value for the grow step from $\beta_i$.
So our rule grows $P_1$ as desired.

Suppose $i$ is even.
The distance of $f$'s end  from the start of $P_1$,
\eqref{Distance_fEndEqn}, is even.
So
$P1$ can be extended to $f$'s end when $2\delta= 3i-2$.
This corresponds to $y(\phi)=-b-1-2\delta=-b-1-(3i-2)= -b-3i+1$.
\eqref{yPhiGoalEqn} gives the same value for the grow step from $\beta_i$.
Again our rule grows $P_1$ as desired.

\bigskip

Edge $e$ triggers a blossom step in
the $f$-matching algorithm.

In summary, 
the uniqueness of the structured matching, plus the above $P_1$ growth argument,
implies
that the $f$-matching algorithm finds the search structure that
we show.

We turn to the structure of the cross edges.
The alternating trail from free vertex
$\beta_b$ to the cross edge
from $\beta_i$ is 3 edges longer than the similar
trail for $\beta_{i-1}$. This is  follows
since the $\beta_i$ trail has 3 extra edges. It
can be constructed from $\beta_{i-1}$ by replacing
$\beta_i\beta_{i-1}s_{i-1,1}$
with $\beta_i s_{i1}s_{i2}s_{i3}s_{i4}$
and extending it with edge $\beta_{i-1}\beta_{i}$.
Fig.\ref{GeneralCrossersFig} illustrates the cross edges
as they switch the path leading to $e$.
The pattern repeats every 3 cross overs.

\def\switch{0}
\ifcase \switch

\or

\input HK
\input time
\input LG
\newpage
\input ATBound
\newpage
\input aLevel
\newpage
\input FdEdge
\newpage
\input GRespect
\newpage

\input code
\fi

\ifcase 1  
\or

\fi

\begin{thebibliography}{99}
\footnotesize

\def\referencesheading{0}
\ifcase\referencesheading
\or
\bigskip\penalty-2000%
\noindent{\bf References \hfill\bigskip}
\parindent=0pt
\or
\line{\quad}\penalty-2000%
\noindent{\twelvebf References \hfill}
\medskip
\parindent=0pt
\fi
%
\font\ninerm=cmr9
\font\ninebf=cmbx9
\font\nineit=cmti9        
\font\ninei=cmmi9
\font\ninesy=cmsy9
\def\ninepoint{%
   \def\rm{\ninerm}\def\bf{\ninebf}%
   \def\it{\nineit}\def\smc{\ninerm}\baselineskip=11pt\rm%
        \textfont0=\ninerm \scriptfont0=\sevenrm
	\textfont1=\ninei \scriptfont1=\seveni
	\textfont2=\ninesy \scriptfont2=\sevensy
	\textfont3=\tenex \scriptfont3=\tenex
}
\def\rsize{39pt}
\def\r #1]{\medskip%
\hangafter=-10\hangindent=\rsize%
\hskip0pt  \llap{\hbox to \rsize{#1]\hfill}}\ignorespaces}
%
%
\def\al #1,{{\it Algorithmica, #1,}}
\def\comb #1,{{\it Combinatorica, #1,}}
\def\cu,{Comp.\ Sci.\ Dept., Univ.\ Colorado, Boulder, CO,}
\def\focs #1,{{\it Proc.\ #1 Annual Symp.\ on Found.\ of Comp.\ Sci.,}}
\def\ipl #1,{{\it Inf.\ Proc.\ Letters, #1,}}
\def\ja #1,{{\it J.\ Algorithms,  #1,}}
\def\jacm #1,{{\it J.\ ACM,  #1,}}
\def\jcss #1,{{\it J.\ Comp.\ and System Sci., #1,}}
\def\mprog #1,{{\it Math.\ Programming,  #1,}}
\def\mprogb #1,{{\it Math.\ Programming B,  #1,}}
\def\net #1,{{\it Networks, #1,}}
\def\phd{Ph.\ D.\ Dissertation}
\def\sicomp #1,{{\it SIAM J.\ Comput.,  #1,}}
\def\siad #1,{{\it SIAM J.\ Alg.\ Disc.\ Meth., #1,}}
\def\sidm #1,{{\it SIAM J.\ Disc.\ Math., #1,}}
\def\soda #1,{{\it Proc.\ #1 Annual ACM-SIAM Symp.\ on Disc.\ Algorithms,}} 
\def\stoc #1,{{\it Proc.\ #1 Annual ACM Symp.\ on Theory of Comp.,}}
\def\tr {Tech. Rept.\ }
%
\def\pp#1-#2.{pp.\ #1--#2.}
\def\spp#1-#2;{pp.\ #1--#2;}

%
\def\nrlq #1,{{\it Naval Res.\ Logist.\ Quart., #1,}}
\def\talg #1,{{\it ACM Trans.~Algorithms}  #1,}

\bibitem{AMO}
R.K. Ahuja, T.L. Magnanti, and J.B. Orlin,
{\em Network Flows: Theory, Algorithms, and Applications},
Prentice-Hall, Saddle River, New Jersey, 1993.

\bibitem{CKLPGS}
  L. Chen, R. Kyng, Y. Liu, R. Peng, M.P. Gutenberg, S. Sachdeva,
  ``Maximum flow and minimum-cost flow in almost-linear time'',
arXiv:2010.01102, 2022.


\bibitem{D}
E.A. Dinic, 
``Algorithm for solution of a problem
of maximum flow in networks with power estimation'', in Russian,
{\em Soviet Mathematics Doklody} 11,
1970, \pp 1277-1280.

\bibitem{DHZ}
R. Duan, H. He, and T. Zhang,
"A scaling algorithm for weighted f-factors in general graphs",
{\em Proc. of the  47th International
Colloquium on Automata, Languages, and Programming (ICALP 2020)},
Vol. 168 of LIPIcs, \pp 41:1-41:17, 2020.

\bibitem{DPS}
R. Duan, S. Pettie, and H-H. Su,
"Scaling algorithms for weighted matching in general graphs",
{\em ACM Trans. Algorithms} 14, 1,
2018,
Article 8, 35 pages.

\bibitem{Ed}
J. Edmonds, 
"Paths, trees, and flowers",
Canad. J. Math. 17, 1965, \pp 449-467.

\bibitem{E}
J. Edmonds, ``Maximum matching and a polyhedron with 0,1-vertices'', 
{\it  J.\ Res.\ Nat.\ Bur.\ Standards 69B}, 1965, \pp 125-130.


\bibitem{EK}
J. Edmonds and R.M. Karp,  
"Theoretical improvements in algorithmic efficiency for network flow problems",
\jacm 19,  1972, \pp 248-264.

\bibitem{ET}
S. Even and R.E. Tarjan,
``Network flow and testing graph connectivity'',
\sicomp 4, 1975, \pp 507-518.

\bibitem{G17}
 H.N.~Gabow,
"The weighted matching approach to maximum cardinality matching,"
{\it Fundamenta Informaticae (Elegant Structures in Computation:
To Andrzej Ehrenfeucht on His 85th Birthday)
154}, 1-4, 2017, \pp 109-130.

\bibitem{G18}
 H.N.~Gabow,
"Data structures for weighted matching and extensions to
$b$-matching and $f$-factors,"
\talg 14, 3, 2018, Article 39, 80 pages.

\bibitem{G21}
H.N.~Gabow,
``Blocking trails for $f$-factors of multigraphs,''
{\it Algorithmica},
https://doi.org/10.1007/s00453-023-01126-y,
2023.

\bibitem{G23b}
  H.N.~Gabow,
"A weight-scaling algorithm for $f$-factors of multigraphs,"
{\it Algorithmica}, \hfill\break
https://doi.org/10.1007/s00453-023-01127-x,
2023.

\bibitem{GT89}
H.N. Gabow and R.E. Tarjan,
``Faster scaling algorithms for network problems,''
\sicomp 18, 5, 1989, \pp 1013-1036.

\bibitem{HK}
J.E. Hopcroft and R.M. Karp,
``An $n^{2.5}$ algorithm for maximum matchings in bipartite
graphs'', \sicomp 2, 1973, \pp 225-231.

\bibitem{HP}
D. Huang and S. Pettie, ``Approximate generalized matching:
$f$-matchings and $f$-edge covers",
{\it Algorithmica} 84, 7, 2022,  \pp 1952-1992.

\bibitem{K}
A.V. Karzanov, 
``On finding maximum flows in network with special structure and some
applications'', in Russian, {\em Math.\ Problems for Production Control} 5,
Moscow State University Press, 1973, \pp 81-94.

\bibitem{K73b}
A.V. Karzanov, 
``An exact estimate of an algorithm for finding a maximum flow,
applied to the problems 'of representatives'``, in Russian,
{\em Voprosy Kibernetiki, Trudy Seminara po Kombinatorno\hskip1pt$\breve{\imath}$ Matematike} 5,
Sovetskoe Radio, Moscow,
1973, \pp 66-70.

\bibitem{KV}
  B. Korte and J. Vygen,
  {\em Combinatorial Optimization: Theory and Algorithms},
3rd Ed., Springer, NY, 2005.

\bibitem{MV}
S. Micali and V.V. Vazirani, "An $O(\sqrt {|v|}\cdot |E|)$ algorithm for finding
maximum matching in general graphs",
\focs 21st, 1980, \pp 17-27.

\bibitem{OA}
J.B. Orlin and R.K. Ahuja,
``New scaling algorithms for the assignment and minimum mean cycle problems'',
{\em Math. Programming} 54, 1992, \pp 41-56. 

\bibitem{S}
A.~Schrijver,
{\it Combinatorial Optimization: Polyhedra and Efficiency},
Springer, NY, 2003.


\bibitem{V1}
V.V.~Vazirani,
``A theory of alternating paths and blossoms for proving 
correctness of the $O(\sqrt V E)$ general graph maximum matching
algorithm'',
{\em Combinatorica}, 14, 1, 1994, \pp 71-109.

\bibitem{V2}
V.V.~Vazirani,
``A simplification of the MV matching algorithm and its proof'',
 {\em CoRR}, abs/1210.4594v5, Aug.27, 2013, 32 pages;
also
``A proof of the MV matching algorithm'', manuscript, May 13, 2014, 42 pages.



\end{thebibliography}
\end{document}